\documentclass[aps, prd, twocolumn, showpacs, superscriptaddress, groupedaddress, floatfix]{revtex4-2}
\usepackage{caption}
\usepackage{graphicx}
\usepackage{subcaption}
\usepackage[dvipsnames]{xcolor}% Include figure files
\usepackage{yfonts}
\usepackage{amsmath}
\usepackage{amssymb}
\usepackage{amsfonts}
\usepackage{amsthm}
\usepackage{mathtools}
\usepackage{mathrsfs}
\usepackage{derivative}
\usepackage{dcolumn}% Align table columns on decimal point
\usepackage{listings}
\usepackage{comment}
\usepackage{rotating, bm, array}
\usepackage[utf8]{inputenc}
\usepackage[english]{babel}
\usepackage[normalem]{ulem}
\usepackage[pagebackref=false, colorlinks=true]{hyperref}
\hypersetup{
linkcolor=blue,     % color of internal links
citecolor=blue,     % color of links to bibliography
urlcolor=blue}      % color of the url

\begin{document}

\title{Junction Conditions and Gravitational Collapse in Scalar-Tensor-Vector Gravity}

\author{Debanjan Debnath}
\email{debjanjan@gmail.com}
\author{Anant Badal}
\email{anantbadal22@iitk.ac.in}
\author{Kaushik Bhattacharya}
\email{kaushikb@iitk.ac.in}
\affiliation{Department of Physics, Indian Institute of Technology Kanpur, Uttar Pradesh - 208016, India}

\begin{abstract}
  We formulate the junction conditions for Scalar-Tensor-Vector Gravity (STVG/MOG), proposed by J.~W.~Moffat. Using these conditions, the theory of gravitational collapse is constructed. In the collapsing process, an interior Friedmann-Lema\^itre-Robertson-Walker (FLRW) spacetime with baryonic matter and dark energy is matched with an exterior static, spherically symmetric Reissner--Nordstr\"{o}m (RN)-like spacetime through a shell that carries STVG-charge. Starting from the standard STVG action, we derive the junction conditions across a boundary that relate the values of the various field quantities and their derivatives across the matching surface. Using the matching conditions and the nature of the collapsing shell, it is shown that a gravitational collapse can proceed in the present situation, and one can have RN-like horizon formation in finite proper time. We present two simplified models of gravitational collapse in this article: one ends up as an extremal RN-like black hole, and the other tends to collapse towards a sub-extremal RN-like black hole, as observed by an asymptotic observer at an infinite distance away from the collapsing system. 
\end{abstract}

\maketitle

%%%%%%%%%%%%%%%%%%%%%%%%%%%%%
\section{Introduction}

Dark matter was introduced as an extra matter component in addition to the baryonic one, in order to explain the so-called missing mass problem first observed by Fritz Zwicky in 1933 \cite{MissingMass}. In this regard particle physicists have advanced many dark matter models, none of which have been verified observationally. There are various interesting particle physics models about dark matter, here we mention some of the relevant literature on this topic; see Refs.~\cite{Arbey:2021gdg,Bertone:2016nfn}. Research along these various particle physics models are not the only directions in which physicists have tried to address the missing mass problem.

It was suggested that gravitational theory itself can be responsible for the missing mass problem in astrophysics. Modified Newtonian dynamics (MOND) \cite{Milgrom:1983ca, Milgrom:2004ba, Famaey:2011kh} tries to address the missing mass problem on galactic scales by modifying Newton's law of gravity. Although MOND tried to tackle the missing mass problem in a novel way, it has one major drawback. MOND is a non-relativistic theory. To tackle this particular problem, Bekenstein has suggested a relativistic generalization of MOND, which is called the Tensor-Vector-Scalar (TeVeS) theory \cite{Bekenstein:2004ne, Famaey:2011kh, Chaichian:2014dfa}. This theory contains not only the standard tensor degree of freedom, as we have in general relativity (GR), but also has vector as well as scalar degrees of freedom.  It has been observed that the original TeVeS theory cannot properly describe various properties of gravitational waves and galaxy clusters \cite{Sagi:2010ei,Gong:2018cgj, Feix:2010pn, Takahashi:2007nj}.     

In this regard there are other modified theories of gravity (MOG) which try to address the missing mass problem by introducing new fields. In Ref.~\cite{Moffat2006} Moffat has shown that there can be a covariant scalar-tensor-vector gravity (STVG) theory where the gravitational constant $G$, a vector field coupling $\omega$ and the vector field $A_\mu$ having mass $\varphi$, all vary with space and time. This theory has the main characteristic that, in addition  to the generic gravitational force which is proportional to the gravitational constant $G$, a Yukawa-type force also comes into play which is repulsive in nature. This repulsive fifth-force is generated by the massive Proca vector field $A_\mu$. The proposed theory becomes important as the equations of motion for a test particle lead to a modified gravitational acceleration law that can fit galaxy rotation curves and cluster data without non-baryonic dark matter \cite{Moffat:2013sja}. The theory is consistent with solar system observational tests. The linear evolutions of the metric, vector field, and scalar field perturbations and their consequences for the observations of the cosmic microwave background are investigated \cite{MoffatCMB, Jamali}. In Ref.~\cite{MoffatStructure}, the author showed that, the role of dark matter in the enhanced growth of structure can also be reproduced by the dominant component of the energy density of the vector field in the early universe. The corresponding matter power spectrum was also shown to be in agreement with the observation. The STVG theory proposed by Moffat appears to be a promising theory and one can find some more interesting applications of this theory in Ref.~\cite{Moffat:2020mug, Brownstein:2007sr, Moffat:2019uxp}.

In the present article we will address the issue of gravitational collapse in the STVG theory. Our aim is to study the gravitational collapse of a spherical region, containing baryonic matter and dark energy, into a black hole. We would like to investigate whether a process analogous to the Oppenheimer-Snyder-Datt collapse \cite{Oppenheimer:1939ue, datt} can occur in STVG theory.  Unlike standard collapse scenarios in GR, where an overdense region of dark matter undergoes gravitational collapse, in STVG theory we concentrate on a spherical patch of baryonic matter and dark energy as STVG theory does not require explicit presence of dark matter. Black hole formation from a gravitational collapse can occur when an internal dynamical spacetime collapses, and for consistency this spacetime is needed to be glued with an exterior black hole spacetime. In our case we will assume an internal FLRW spacetime and an external black hole spacetime which is like the Reissner--Nordstr\"{o}m spacetime. One can only formulate the collapse process if one knows the junction conditions in the STVG theory. In this paper, we first present the junction conditions in the STVG theory and then apply it for gluing the relevant spacetimes.   

In Ref.~\cite{Moffat2006} the author presented two basic solutions of STVG theory, one solution was like the Reissner--Nordstr\"{o}m (RN) solution and the other one was the FLRW solution in the presence of matter. We call the spherically symmetric spacetime RN-like spacetime as in STVG theory there is no electric charge present and the external 4-vector is actually a component of the gravitational sector. Outside the RN-like outer horizon one can have the zeroth component of the 4-vector part of STVG theory which depends only on the radial coordinate. The homogeneous and isotropic FLRW spacetime can accommodate only the zeroth component of the 4-vector, which can depend only on time. In the present work, it is shown that the RN-like external spacetime can be matched with an FLRW inner collapsing core through an STVG-charged shell which acts as the source of the zeroth component of the external 4-vector.  

In the gravitational collapse we study, the junction conditions demand that the internal FLRW spacetime is matched with the RN-like external spacetime via a STVG-charged shell. The primary reason for it is that the zeroth component of the 4-vector from the two spacetime regions do not match continuously at the junction and hence there is a discontinuity of the field strength tensor along the junction. The discontinuity of the field strength tensor requires an STVG-charge density on the shell separating the two spacetimes along the junction. This charge density acts as the source of the zeroth component of the 4-vector present in the external RN-like spacetime. One can generally assume that the zeroth component of the 4-vector field to be zero in the internal FLRW spacetime. 

We have analyzed the gravitational collapse process in terms of the proper time of an observer participating in the collapse and during the gravitational collapse the junction with the STVG-charged shell can cross the event horizon (null outer Killing horizon) of the RN-like spacetime and ultimately approach the null inner Killing horizon which also happens to be a Cauchy horizon. It is known that collapsing matter reaching the inner Cauchy horizon is subjected to instabilities \cite{Poisson:1989zz, Ori:1991zz, mihalis} and the energy-momentum tensor of the collapsing shell junction may show irregular behavior. In our case the shell energy-momentum is well-behaved as long as the shell is outside the Cauchy horizon of the RN spacetime. As the shell tends towards the second horizon the shell energy density starts to decrease rapidly showing instability. In another case we have shown how a gravitational collapse can produce an extremal RN-like black hole. The asymptotic observer only observes the STVG-charged shell to move towards the horizon and freeze and fade there.  

Throughout this paper, we work with a system of units where $c = 1$ and with the mostly-plus metric signature $(-,+,+,+)$.
We follow the definitions
\begin{align}
    \Gamma^{\lambda}{}_{\mu\nu} = \frac{1}{2}g^{\lambda\sigma}\left(\partial_{\nu}g_{\sigma\mu} + \partial_{\mu} g_{\sigma\nu} - \partial_{\sigma}g_{\mu\nu}\right),
    \label{christoffel}
\end{align}
and
\begin{align}
    {R^{\rho}}_{\sigma\mu\nu} &= \partial_\mu{\Gamma^{\rho}}_{\sigma\nu} - \partial_\nu {\Gamma^{\rho}}_{\sigma\mu} + \Gamma^{\rho}{}_{\mu\lambda}\Gamma^{\lambda}{}_{\nu\sigma} - \Gamma^{\rho}{}_{\nu\lambda}\Gamma^{\lambda}{}_{\mu\sigma},\nonumber\\
    R_{\mu\nu} &\equiv {R^\rho}_{\mu\rho\nu}, \,\, R \equiv g^{\mu\nu}R_{\mu\nu}.
    \label{riemann}
\end{align}
In addition, $g$ denotes the determinant constructed out of the metric components $g_{\mu\nu}$, and $\nabla_\mu$ denotes the Levi-Civita covariant derivative associated with $g_{\mu\nu}$. While we treat the ``gravitational constant'' $G$ to be a scalar field, we have kept the cosmological constant $\Lambda$ strictly as a constant.

The material in this article is presented in the following way. In the next section we present the basic field equations of STVG theory and specify our notation and convention by writing down the basic equations of the STVG theory. In Sec.~\ref{junconds} we derive the junction conditions of the STVG theory. Section~\ref{sptimes} solves the field equations in the interior and exterior spacetimes involved in the collapsing process. In Sec~\ref{NumSol} we present the collapsing solutions in two simplified STVG collapsing models. The last section summarizes the main points presented in the article. 

%%%%%%%%%%%%%%%%%%%%%%%%%%%%%%%%%%%%%%%%%%%%%%%%%%%%%%%%%%%%%%%%%%%%%%%%%%%%%%%%%%%%%%%%%%%%%%%%%%%
\section{Field Equations and Equations of Motion}
\label{sec:field_eqs_eoms}

In this section, we present the relevant equations of STVG theory; these equations were first derived in Ref.~\cite{Moffat2006}. The total action is taken to be
\begin{align}
    \mathcal{S} = \mathcal{S}_{\text{grav.}} + \mathcal{S}_{V} + \mathcal{S}_{S} + \mathcal{S}_{M},
\end{align}
where $\mathcal{S}_{\text{grav.}}$, $\mathcal{S}_{V}$, $\mathcal{S}_{S}$ and $\mathcal{S}_{M}$ correspond respectively to the gravitational, vector, scalar and matter sectors. In our convention:
\begin{align}
    \mathcal{S}_{\text{grav.}}
    &=
    \frac{1}{16\pi}
    \int d^4x\,\sqrt{-g}\,
    \frac{1}{G}\left(R-2\Lambda\right),
    \label{SEH}
\end{align}
\begin{align}
    \mathcal{S}_{V} &= -\int d^4x\,\sqrt{-g}\,\omega\left[\frac{1}{4} B^{\mu\nu}B_{\mu\nu} + V_A(A_\mu)\right],
    \label{SV}
\end{align}
and
%\begin{align}
%    \mathcal{S}_{S} =
%    \int d^4x\,\sqrt{-g}\,\Bigg[&\frac{1}{G^3}\left(-\frac{1}{2}\nabla_\mu G\nabla^\mu G - V_G(G)\right)\nonumber\\
%    &+ \frac{1}{G}\left(-\frac{1}{2}\nabla_\mu \omega \nabla^\mu \omega
%    - V_\omega(\omega)\right)\nonumber\\
%    &+ \frac{1}{\varphi^2 G}\left(-\frac{1}{2}\nabla_\mu \varphi \nabla^\mu \varphi - V_\varphi(\varphi)\right)\Bigg].
%    \label{SS}
%\end{align}
\begin{align}
    \mathcal{S}_{S} =
    \int d^4x\,\sqrt{-g}\, &\left\{\frac{1}{G^3}\left(-\frac{1}{2}\nabla_\mu G\nabla^\mu G - V_G(G)\right) \right. \nonumber\\
    & \left.+ \frac{1}{G}\left(-\frac{1}{2}\nabla_\mu \omega \nabla^\mu \omega
    - V_\omega(\omega)\right) \right. \nonumber\\
    &\left. + \frac{1}{\varphi^2 G}\left(-\frac{1}{2}\nabla_\mu \varphi \nabla^\mu \varphi - V_\varphi(\varphi)\right)\right\}.
    \label{SS}
\end{align}
Here $A_\mu$ is the massive vector field, whose mass is promoted to the scalar field $\varphi(x)$. The fields $G(x)$ and $\omega(x)$ are respectively the gravitational coupling and the vector-field coupling, and are also treated as scalar fields. The terms $V_A(A_\mu)$, $V_G(G)$, $V_\varphi(\varphi)$, and $V_\omega(\omega)$ represent the corresponding field potentials in STVG theory.

The field strength is defined by
\begin{align}
    B_{\mu\nu}
    \equiv
    \nabla_\mu A_\nu - \nabla_\nu A_\mu
    =
    \partial_\mu A_\nu - \partial_\nu A_\mu.
    \label{Bdef}
\end{align}
Varying the total action with respect to $g^{\mu \nu}$ we obtain the field equations:
\begin{align}
    G_{\mu \nu} + \Lambda g_{\mu \nu} + Q_{\mu \nu} = 8 \pi G T_{\mu \nu}, \label{FE}
\end{align}
where
\begin{subequations}
\label{eq:main_label}
\begin{equation}
    Q_{\mu\nu} \equiv G\left(g_{\mu\nu}\nabla^\lambda\nabla_\lambda G^{-1} - \nabla_\mu\nabla_\nu G^{-1}\right)\label{Qdef}
\end{equation}
\begin{align}
    &\qquad \quad = g_{\mu\nu}\left(-\frac{1}{G}\nabla^\rho\nabla_\rho G + \frac{2}{G^2}\nabla^\rho G \,\nabla_\rho G \right) \nonumber \\
    &\qquad\qquad + \frac{1}{G} \nabla_\mu \nabla_\nu G - \frac{2}{G^2}  \nabla_\mu G \, \nabla_\nu G. \label{Qexp}%\cite{Milgrom:1983ca}
\end{align}
\end{subequations}
The total stress tensor $T_{\mu\nu}$ is given by
\begin{align}
    T_{\mu \nu} = \prescript{(M)}{}T_{\mu \nu} + \prescript{(V)}{}T_{\mu \nu} + \prescript{(S)}{}T_{\mu \nu},
\end{align}
where $\prescript{(M)}{}T_{\mu\nu}$, $\prescript{(V)}{}T_{\mu \nu}$ and $\prescript{(S)}{}T_{\mu \nu}$ are the stress tensors corresponding to the matter sector, vector part and the scalar sector respectively, with $\prescript{(S)}{}T_{\mu\nu} = \prescript{(G)}{}T_{\mu\nu} + \prescript{(\omega)}{}T_{\mu\nu} + \prescript{(\varphi)}{}T_{\mu\nu}$ and these are given by a common defining equation
\begin{equation}
    \prescript{(X)}{}T_{\mu\nu} = -\frac{2}{\sqrt{-g}}\frac{\delta \mathcal{S}_{X}}{\delta g^{\mu\nu}}, \qquad X \in \{M, V, G, \omega, \varphi\}.
\end{equation}%\cite{Milgrom:1983ca}
Expanding, we obtain
\begin{widetext}
\begin{align}
    \prescript{(V)}{}T_{\mu\nu} &= \omega \left\{B_{\mu}{}^{\lambda}B_{\nu\lambda} - g_{\mu\nu} \left(\frac{1}{4}B^{\sigma\rho}B_{\sigma\rho} + V_A(A_\mu)\right) + 2\frac{\partial V_A(A_\mu)}{\partial g^{\mu\nu}}\right\},\label{SETA}\\
    \prescript{(G)}{}T_{\mu \nu} &= - \frac{1}{G^3} \left\{\nabla_\mu G \nabla_\nu G -  2 \frac{\partial V_G (G)}{\partial g^{\mu \nu}}  -  g_{\mu \nu} \left( \frac{1}{2}  \nabla_\lambda G \nabla^\lambda G - V_G \left( G \right)\right)\right\}, \label{SETG} \\
     \prescript{(\omega)}{}T_{\mu \nu} &= - \frac{1}{G} \left\{\nabla_\mu \omega \nabla_\nu \omega -  2 \frac{\partial V_\omega (\omega)}{\partial g^{\mu \nu}}  -  g_{\mu \nu} \left( \frac{1}{2}  \nabla_\lambda \omega \nabla^\lambda \omega - V_\omega \left( \omega \right)\right)\right\}, \label{SETomega} \\
     \prescript{(\varphi)}{}T_{\mu \nu} &= - \frac{1}{\varphi^2 G} \left\{\nabla_\mu \varphi \nabla_\nu \varphi -  2 \frac{\partial V_\varphi (\varphi)}{\partial g^{\mu \nu}}  -  g_{\mu \nu} \left( \frac{1}{2}  \nabla_\lambda \varphi \nabla^\lambda \varphi - V_\varphi \left( \varphi \right)\right)\right\}. \label{SETphi}
\end{align}
\end{widetext}
Taking the variation of the total action with respect to $A_{\mu}$, we get: 
\begin{align}
    \nabla_{\nu} B^{\mu \nu} + \frac{\partial V_A (A_\mu)}{\partial A_{\mu}} + \frac{1}{\omega}  B^{\mu \nu}\nabla_{\nu} \omega = - \frac{1}{\omega} J^{\mu}, 
    \label{EoM_A}
\end{align}
where
\begin{align}
    \frac{1}{\sqrt{- g}} \frac{\delta \mathcal{S}_M}{\delta A_\mu} \equiv - J^{\mu}.
\end{align}
We also assume conservation of the matter current:
\begin{align}
    \nabla_\mu J^\mu = 0. \label{divJZero}
\end{align}
From Eq.~(\ref{EoM_A}) and the above equation we see that $J^\mu$ acts as a conserved source current for the 4-vector field $A^\mu$ in the theory and one can define a STVG-charge by appropriately integrating $J^0$ on the spatial part of the spacetime manifold. By taking the four-divergence of Eq.~\eqref{EoM_A} and using Eq.~\eqref{divJZero}, we get
\begin{align}
    \nabla_\mu \nabla_{\nu} B^{\mu \nu} &+ \nabla_\mu \left(\frac{\partial V_A (A_\mu)}{\partial A_{\mu}} \right) \nonumber \\ &+ \nabla_\mu \left(\frac{1}{\omega}  B^{\mu \nu}\nabla_{\nu} \omega \right) = - J^\mu \nabla_\mu \left(\frac{1}{\omega} \right). \label{divEoMA}
\end{align}
%For future reference, we note that, the first term of Eq.~\eqref{divEoMA} can be rewritten as 
%\begin{align}
%    \nabla_\mu \nabla_\nu B^{\mu \nu} = \partial_\mu &\left\{\frac{1}{\sqrt{- g}} \partial_\nu \left(\sqrt{- g} B^{\mu \nu}   
%\right) \right\} \nonumber \\ &+ \frac{1}{\sqrt{-g}} \partial_\mu \left(\sqrt{-g} \right) \partial_\nu B^{\mu \nu},
%\end{align}
%and can be shown to be zero. The above relation is true for any second-rank antisymmetric tensor $B_{\mu\nu}$.\\
By varying the total action with respect to $G$, we get:
\begin{align}
   \nabla_{\mu}\nabla^{\mu}G - V_G'(G) + \mathscr{N} = -\frac{1}{2}G^2\left(T - \frac{\Lambda}{4\pi G}\right), \label{EoM_G}
\end{align}
where $T = g^{\mu\nu}T_{\mu\nu}$, and
\begin{widetext}
    \begin{align}
    \mathscr{N} &= -\frac{3G^2}{16\pi}\nabla_{\mu}\nabla^{\mu}G^{-1} - \frac{3}{G}\left(\frac{1}{2}\nabla_{\mu}G\nabla^{\mu}G - V_G(G)\right) + G\left(\frac{1}{2}\nabla_{\mu}\omega\nabla^{\mu}\omega + V_{\omega}(\omega)\right) + \frac{G}{\varphi^2}\left(\frac{1}{2}\nabla_{\mu}\varphi\nabla^{\mu}\varphi + V_{\varphi}(\varphi)\right). \label{N}
\end{align}
\end{widetext}
%We note that we can rewrite the first term in the expression for $\mathscr N$ as
%\begin{align}
%    \frac{3}{16 \pi} \nabla^\mu \nabla_\mu G - \frac{3}{8 \pi} \frac{1}{G} \nabla^\mu G \nabla_\mu G.
%\end{align}
Varying $\omega$, we obtain its equation of motion:
\begin{align}
   \nabla^\mu\nabla_\mu \omega - V_\omega'(\omega) - \mathscr{F} = 0, \label{EoM_omega}
\end{align}
where
\begin{align}
    \mathscr{F} \equiv G^{-1}\nabla_\mu G \nabla^\mu \omega + G\left\{
    \frac{1}{4}B_{\mu\nu}B^{\mu\nu} + V_A(A_\mu)\right\}. \label{F}
\end{align}
Similarly, for the field $\varphi$, we have
\begin{align}
    \nabla_\mu \nabla^\mu \varphi - V_\varphi^\prime (\varphi) + \mathscr{P} = 0, \label{EoM_phi}
\end{align}
where
\begin{align}
    \mathscr{P} \equiv &-\left\{G^{-1}\nabla_\mu G \nabla^\mu \varphi + \frac{1}{\varphi}\nabla_\mu \varphi \nabla^\mu \varphi \right.\nonumber\\
    &\qquad \left.- \frac{2}{\varphi}V_\varphi(\varphi) +  \omega G \varphi^2 \frac{\partial V_A(A_\mu)}{\partial \varphi}\right\} \label{P}
\end{align}
%The above equation does not exactly match with the corresponding equation [Eq.~(27)] given in Ref.~\cite{Moffat2006}. 
These are the relevant field equations in the STVG theory we will be using in this article.
%%%%%%%%%%%%%%%%%%%%%%%%%%%%%%%%%%%%%%%%%%%%%%%%%%%%%%%%%%%%%%%%%%%%%%%%%%%%%%%%%%%%%%%%%
\section{Calculation of the Junction Conditions for the Field Variables}
\label{junconds}

To calculate the junction conditions, we will follow the symbols and notations of E. Poisson [Sec. (3.7) of Ref.~\cite{PoissonToolkit}]. The hypersurface $\Sigma$ separates spacetime into two regions, $\mathscr{V}^+$ and $\mathscr{V}^-$, within which the spacetimes are described by the  metrics $g^{+}_{\mu \nu}$ and $g_{\mu \nu}^-$, which are expressed in coordinates $x^\mu_+$  and $x^\mu_-$  respectively. The objective is to obtain the constraint equations (matching conditions) which we must impose on the various geometrical quantities, e.g., metric and other gravitational degrees of freedom, such that, the regions $\mathscr{V}^+$ and $\mathscr{V}^-$ are joined smoothly on $\Sigma$.

On the matching-hypersurface $\Sigma$, we set the coordinates \{$y^a$\} where $a$ runs from $1$ to $3$, and $y^a \equiv \{\tau, \vartheta, \phi \}$. The tangent vectors to the hypersurface are $e^\mu_a := \left(\frac{\partial x^\mu}{\partial y^a} \right)$. We use $\ell (x^\mu)$ as proper length or proper time along the geodesics which intersect the hypersurface orthogonally, so that the unit normal to the hypersurface can be written as: $n_{\mu} = \epsilon \partial_{\mu}\ell$ and consequently $dx^\mu = n^\mu d\ell $. Here $\epsilon \equiv n^\mu n_\mu$ is a constant which is $-1$ or $+1$ for spacelike or timelike hypersurface respectively. We can check that, $n_\mu e^\mu_a = 0$ by construction.

In this theory, there are  five fields altogether: the metric $g_{\mu \nu}$, the vector field $A_{\mu}$ and the three scalar fields $G$, $\omega$ and $\varphi$. We will consider the behavior of these fields and their derivatives across the hypersurface
$\Sigma$. For the vector field $A_\mu$ we will match the field strength tensor $B_{\mu \nu}$  on the junction, a direct matching of the 4-vector field does not give any interesting result. To maintain clarity, we will calculate each of these conditions corresponding to the five fields in separate subsections.  

%%%%%%%%%%%%%%%%%%%%%%%%%%%%%%%%%%%%%%%%%%%%%%%%%%%%%%%%%%%%%%%%%%%%%%%%%%%%%%%%%%%%%%%%%%%%%%%%%%%%%%%%%%%%%%%%%%%%%%%%%%%%
\subsection{Junction Conditions for the Metric Field}

 We write the metric components as 
\begin{align}
    g_{\mu \nu} = \Theta (\ell) g_{\mu \nu}^+ + \Theta (-\ell) g_{\mu \nu}^- , 
\end{align}
and consequently
\begin{align}
    \partial_\lambda g_{\mu \nu} = \Theta (\ell) \, \partial_\lambda g_{\mu \nu}^+ + \Theta (-\ell) \, \partial_\lambda g_{\mu \nu}^- + \epsilon \delta (\ell) \left[ g_{\mu \nu} \right] n_\lambda,
\end{align}
where for any quantity $X$, we have defined
\begin{align}
    \left[X \right] \equiv  X^+ - X^-.
\end{align}
From Eq.~\eqref{christoffel}, we can see that to avoid terms like $\Theta (\ell) \delta (\ell)$, which is not defined in a distributional sense, we need to impose
\begin{align}
    \left[ g_{\mu \nu} \right] = 0 \implies g^+_{\mu \nu} = g^-_{\mu \nu}.
\end{align}
So, the derivative of the metric field and the corresponding Christoffel symbols calculated from it are
\begin{align}
   \partial_\lambda g_{\mu \nu} = \Theta (\ell) \, \partial_\lambda g_{\mu \nu}^+ + \Theta (-\ell) \, \partial_\lambda g_{\mu \nu}^-, 
\end{align}
and 
\begin{align}
     \Gamma^\lambda_{ \mu \nu} = \Theta (\ell) \, \Gamma^{+\lambda}_{\mu \nu} + \Theta (-\ell) \, \Gamma^{-\lambda}_{\mu \nu}. \label{christoffel_dist}
\end{align}
Taking the derivative of the above equation, we have
\begin{align}
     \partial_\lambda\Gamma^\rho_{\mu \nu} = \Theta (\ell) \, \partial_\lambda\Gamma^{+\rho}_{\mu \nu} + \Theta (-\ell) \, &\partial_\lambda\Gamma^{-\rho}_{\mu \nu} \nonumber \\  &+ \epsilon \, \delta (\ell) \,\left[\Gamma^\rho_{ \mu\nu} \right] \,n_\lambda.
\end{align}
Using this relation in the first of Eq.~\eqref{riemann}, we have the following distributional expression for the Riemann tensor:
\begin{align}
    R^{\rho}_{\,\, \sigma \mu \nu} = \Theta (\ell) R^{+\rho}_{\,\, \sigma \mu \nu} &+ \Theta (-\ell) R^{-\rho}_{\,\, \sigma \mu \nu} \nonumber \\ &+ \epsilon \, \delta (\ell) \, \left( n_\mu \left[\Gamma^{\rho}_{\,\, \sigma \nu} \right] - n_\nu \left[\Gamma^{\rho}_{\,\, \sigma \mu} \right]  \right).
\end{align}
In the following, we will denote the $\delta$-part of the Riemann tensor as
\begin{align}
    A^{\rho}_{\,\, \sigma \mu \nu} \equiv \epsilon \left(n_\mu \left[\Gamma^{\rho}_{\,\, \sigma \nu} \right] - n_\nu \left[\Gamma^{\rho}_{\,\, \sigma \mu} \right] \right).
\end{align}
Taking the trace of the above relation  we get
\begin{align}
    R_{\mu \nu} = \Theta (\ell) R^+_{ \mu \nu} + \Theta (-\ell) R^{-}_{ \mu \nu} +  \, \delta (\ell) \, A_{\mu \nu}.
\end{align}
Contracting this relation with $g^{\mu \nu}$, we get
\begin{align}
    R = \Theta (\ell) \, R^+ + \Theta (-\ell) \, R^{-} +  \delta (\ell) \, A.
\end{align}
In the above two expressions
\begin{align}
    &A_{\mu \nu} \equiv A^{\rho}_{\,\, \mu \rho \nu} = \epsilon \left(  n_\rho \left[\Gamma^{\rho}_{\,\, \mu\nu} \right] - n_\nu \left[\Gamma^{\rho}_{\,\, \mu \rho} \right] \right), \nonumber \\ 
    &A = g^{\mu \nu} A_{\mu \nu} = \epsilon \left( n_\rho \left[\Gamma^{\rho}_{\,\, \mu\nu} \right] g^{\mu \nu} - n^\mu \left[\Gamma^{\rho}_{\,\, \mu \rho} \right] \right). 
\end{align}
Using the expressions for $R_{\mu \nu}$ and $R$, we can calculate the $\delta$-part of the Einstein tensor $G_{\mu\nu} \equiv R_{\mu \nu} - (1/2) R g_{\mu \nu} $, which is
\begin{align}
    A_{\mu \nu} - \frac{1}{2} A g_{\mu \nu}. \label{ETdelta}
\end{align}
We can relate this to the $\delta$-part of the stress-energy tensor $T_{\mu \nu}$. From the relations derived in the next section, it can be shown that the singular part of $Q_{\mu \nu}$ is $\epsilon G^{-1} \left( - g_{\mu \nu} n^\lambda \left[\partial_\lambda G \right] + n_\mu \left[\partial_\nu G \right] \right) $; which combined with the above result yield the singular part of the field equation (Eq.~\eqref{FE}) as
\begin{align}
    A_{\mu \nu} - \frac{1}{2} A g_{\mu \nu} + \epsilon G^{-1} \left( - g_{\mu \nu} n^\lambda \left[\partial_\lambda G \right] + n_\mu \left[\partial_\nu G \right] \right) \nonumber \\ = 8 \pi G S_{\mu \nu}. \label{FEdelta}
\end{align}
To write the above relation we have assumed that the total stress-energy tensor has the following form:
\begin{align}
    T_{\mu \nu} = \Theta (\ell) \, T^+_{\mu \nu} + \Theta (- \ell) \, T^-_{\mu \nu} + \delta (\ell) \, S_{\mu \nu}. \label{SETdecomp}
\end{align}
The above result shows that there can be a shell at the junction which carries its own energy-momentum tensor.

At this point, it can be understood that, the tensor $S_{\mu \nu}$, which corresponds to the matter distribution on the hypersurface $\Sigma$, is by assumption a symmetric tensor.  The $\delta$-part (Eq.~\eqref{ETdelta}) which comes from the Einstein tensor $G_{\mu \nu}$ is also a symmetric tensor. So, for consistency, the $\delta$-part arising from $Q_{\mu \nu}$ must also be symmetric in it's indices. This compels us to equate the expression in the parenthesis of Eq.~\eqref{FEdelta} to zero. So we write
\begin{align}
    - g_{\mu \nu} n^\lambda \left[\partial_\lambda G \right] + n_\mu \left[\partial_\nu G \right] = 0. \label{Qcond}
\end{align}
Eq.~\eqref{Qcond} will be satisfied, if we assume
\begin{align}
    \left[\partial_\mu G  \right] = 0. \label{1Gcond}
\end{align}
Implementing this, Eq.~\eqref{FEdelta} becomes
\begin{align}
    A_{\mu \nu} - \frac{1}{2} A g_{\mu \nu}  = 8 \pi G S_{\mu \nu}. \label{FEdelta2}
\end{align}
%From the LHS of Eq.~\eqref{FEdelta}, we can see that, due to the third term, the tensor $S_{\mu \nu}$ is not purely tangential to the hypersurface $\Sigma$, i.e., $n^\mu S_{\mu \nu}$ does not vanish; although it can be checked, that, $n^\nu S_{\mu \nu}$ does. In this context, we note that
From the expressions of $A_{\mu \nu}$ and $A$, we can show by explicit calculations, that
\begin{align*}
    n^\mu \left(A_{\mu \nu} - \frac{1}{2} A g_{\mu \nu} \right) = 0 =  \left(A_{\mu \nu} - \frac{1}{2} A g_{\mu \nu} \right) n^\nu.
\end{align*}
Consequently, $S_{\mu \nu} n^\mu $ and $ S_{\mu \nu} n^\nu$ both vanish. So the tensor $S_{\mu \nu}$ is tangential to the hypersurface $\Sigma$ and the condition in Eq.~\eqref{FEdelta2} can be rewritten as a condition pertaining to the quantities which belong solely to $\Sigma$. Taking the inner product of Eq.~\eqref{FEdelta2} with $e^\mu_a e^\nu_b$, we arrive at the following equation \cite{PoissonToolkit, Israel1966}
\begin{align}
    S_{ab} = - \frac{\epsilon}{8 \pi G} \left( \left[K_{ab} \right] -  \left[K \right] h_{ab} \right). \label{2ndJC}
\end{align}
For a smooth transition across $\Sigma$, we require $S_{ab}$ to be zero. Hence we impose the condition, that, the extrinsic curvature is same on both  sides of $\Sigma$, i.e., 
\begin{align}
    \left[K_{ab} \right] = 0.
\end{align}
This is known as the \textit{second junction condition}. If $S_{\mu \nu}$ vanishes, then taking trace of Eq.~\eqref{FEdelta} we get
\begin{align}
    A = 0.
\end{align}

%%%%%%%%%%%%%%%%%%%%%%%%%%%%%%%%%%%%%%%%%%%%%%%%%%%%%%%%%%%%%%%%%%%%%%%%%%%%%%%%%%%%%%%%%%%%%%%%%%%%%%%%%%%%%%%%%%%%%%%%
\subsection{Junction Conditions for the Scalar Fields} \label{JCscalar}

Let us first write the scalar field in the following distributional form:
\begin{align}
    G = \Theta (\ell) \, G^+ + \Theta (- \ell) \, G^-.
\end{align}
Taking the derivative we get
\begin{align}
    \partial_\mu G = \Theta (\ell) \, \partial_\mu G^+ + \Theta (- \ell) \, \partial_\mu G^- + \epsilon \, \delta (\ell) \left[G \right] n_\mu.
\end{align}
From the equation of motion of $G$ (Eq.~\eqref{EoM_G}), we can see from second term of $\mathscr{N}$, that, we have to impose
\begin{align}
    \left[G \right] = 0 \implies G^+ = G^-,
\end{align}
or there will be terms involving $\Theta (\ell) \delta (\ell)$ in the equation of motion, which are not well defined as distributions. So we rewrite 
\begin{align}
    \partial_\mu G = \Theta (\ell) \, \partial_\mu G^+ + \Theta (- \ell) \, \partial_\mu G^-.
\end{align}
Again taking the derivative we arrive at
\begin{align}
   \partial_\nu \partial_\mu G = \Theta (\ell) \, \partial_\nu \partial_\mu G^+ + \Theta (- \ell) \, &\partial_\nu \partial_\mu G^- \nonumber \\ &+ \epsilon \, \delta (\ell) \left[\partial_\mu G \right] n_\nu.
\end{align}
Using the relation
\begin{align}
    \nabla_\mu \nabla_\nu G = \partial_\mu \partial_\nu G - \Gamma^\lambda_{\,\, \mu \nu} \partial_\lambda G,
\end{align}
which is valid for any scalar field, we get
\begin{align}
    \nabla_\mu \nabla_\nu G = \Theta (\ell) \nabla_\mu \nabla_\nu G^+ + \Theta (- \ell) &\nabla_\mu \nabla_\nu G^- \nonumber \\ &+  \epsilon \, \delta (\ell) \left[\partial_\nu G \right] n_\mu.
\end{align}
Contracting this with $g^{\mu \nu}$, we get
\begin{align}
    \nabla^\mu \nabla_\mu G = \Theta (\ell) \nabla^\mu \nabla_\mu G^+ + \Theta (- \ell) &\nabla^\mu \nabla_\mu G^- \nonumber \\ &+ \epsilon \, \delta (\ell) \left[\partial_\mu G \right] n^\mu.
\end{align}
From the expression of $\mathscr{N}$ (Eq.~\eqref{N}), we can see that, we also need to consider the distributional forms for $\omega$ and $\varphi$. So we assume
\begin{align}
    \omega = \Theta (\ell) \, \omega^+ + \Theta (- \ell) \, \omega^-, \nonumber \\ \text{and} \,\,\, \varphi = \Theta (\ell) \, \varphi^+ + \Theta (-\ell) \, \varphi^-.
\end{align}
Taking the derivative, we have 
\begin{align}
    \partial_\mu \omega = \Theta (\ell) \, \partial_\mu \omega^+ + \Theta (- \ell) \, \partial_\mu \omega^- + \epsilon \,\delta (\ell) \left[\omega\right] \, n_\mu, 
\end{align}
    and
\begin{align}
  \partial_\mu \varphi = \Theta (\ell) \, \partial_\mu \varphi^+ + \Theta (-\ell) \, \partial_\mu \varphi^- + \epsilon \delta (\ell) \left[ \varphi \right] \, n_\mu.
\end{align}
By observing the terms in the third and the fourth parentheses in the expression of $\mathscr{N}$, we conclude that, the $\delta$-terms in both the above equations have to be zero, or there will be cross terms of $\Theta$ and $\delta$, i.e.,
\begin{align}
    \left[\omega\right] = 0 \implies \omega^+ = \omega^-, \,\, \text{and} \,\, \left[\varphi \right] = 0 \implies \varphi^+ = \varphi^-.
\end{align}
We write for $\omega$ and $\varphi$: 
\begin{align}
    \partial_\mu \omega = \Theta (\ell) \, \partial_\mu \omega^+ + \Theta (- \ell) \, \partial_\mu \omega^-, 
\nonumber \\ \text{and} \,\,\,
  \partial_\mu \varphi = \Theta (\ell) \, \partial_\mu \varphi^+ + \Theta (-\ell) \, \partial_\mu \varphi^- .
\end{align}
%Taking the derivative, we get
%\begin{align}
%      \partial_\mu \partial_\nu \omega = \Theta (\ell) \, \partial_\mu \partial_\nu\omega^+ + \Theta (- \ell) \, &\partial_\mu 
%\partial_\nu \omega^-  \nonumber \\ &+ \epsilon \delta (\ell) \left[\partial_\nu \omega \right] n_\mu,
%\end{align}
%and
%\begin{align}
%      \partial_\mu \partial_\nu \varphi = \Theta (\ell) \, \partial_\mu \partial_\nu\varphi^+ + \Theta (- \ell) \, %&\partial_\mu \partial_\nu \varphi^- \nonumber \\ &+ \epsilon \delta (\ell) \left[\partial_\nu \varphi \right] n_\mu. 
%\end{align}
Using these we get
\begin{align}
    \nabla_\mu \nabla_\nu \omega = \Theta (\ell) \nabla_\mu \nabla_\nu \omega^+ + \Theta (- \ell) &\nabla_\mu \nabla_\nu \omega^- \nonumber \\ &+  \epsilon \, \delta (\ell) \left[\partial_\nu \omega \right] n_\mu.
\end{align}
%Contracting this with $g^{\mu \nu}$, we get
%\begin{align}
%    \nabla^\mu \nabla_\mu \omega = \Theta (\ell) \nabla^\mu \nabla_\mu \omega^+ + \Theta (- \ell) \nabla^\mu \nabla_\mu \omega^- + \epsilon \, \delta (\ell) \left[\partial_\mu \omega \right] n^\mu.
%\end{align}
For $\varphi$ we get an exactly similar result.
%Exactly similarly for the field $\varphi$, we get 
%\begin{align}
%    \nabla_\mu \nabla_\nu \varphi = \Theta (\ell) \nabla_\mu \nabla_\nu \varphi^+ + \Theta (- \ell) &\nabla_\mu \nabla_\nu 
%\varphi^-  \nonumber \\ &+  \epsilon \, \delta (\ell) \left[\partial_\nu \varphi \right] n_\mu.
%\end{align}
%Contracting this with $g^{\mu \nu}$, we get
%\begin{align}
%    \nabla^\mu \nabla_\mu \varphi = \Theta (\ell) \nabla^\mu \nabla_\mu \varphi^+ + \Theta (- \ell) \nabla^\mu \nabla_\mu \varphi^- + \epsilon \, \delta (\ell) \left[\partial_\mu \varphi \right] n^\mu.
%\end{align}
As the second and third terms in the equation of motion of the scalar field $\omega$ do not contain any $\delta$-part, we impose 
\begin{align}
     \left[\partial_\mu \omega \right] = 0.
\end{align}
By exactly similar arguments, from the equation of motion of $\varphi$ (Eq.~\eqref{EoM_phi}), we have
\begin{align}
    \left[\partial_\mu \varphi \right] = 0.
\end{align}
Now using the assumed form of stress-energy tensor (Eq.~\eqref{SETdecomp}), and equating  the $\delta$-part of both sides from the equation of motion of $G$ (Eq.~\eqref{EoM_G}), we get 
\begin{align}
  \epsilon  \left(1 + \frac{3}{16 \pi} \right) \left[\partial_\mu G \right] n^\mu = -\frac{1}{2} G^2 S.
   \label{Gcond}
\end{align}
From Eq.~\eqref{1Gcond} we see that $S$, which is the trace of the $\delta$-part of the stress-energy tensor, i.e., $S = g^{\mu \nu} S_{\mu\nu}$ has to vanish. Consequently, if there is a shell at the junction then its energy-momentum tensor has to be traceless in STVG theory, i.e., 
\begin{eqnarray}
S= g^{\mu \nu} S_{\mu\nu} =0\,.
\label{straceless}
\end{eqnarray}
This is an important constraint in the present theoretical model.
%\begin{align}
%    n^\mu \left[\partial_\mu G \right] = 0,
%\end{align}
%which can be seen to be automatically satisfied by the aid of Eq.~\eqref{1Gcond}.
%All these conditions imply, that, the discontinuity of the derivative of the scalar fields have components only along the tangential directions. For simplicity, we can assume that, the discontinuity of the derivative itself vanishes, i.e., 
%\begin{align}
%    \left[\partial_\mu G \right] = 0, \,\,\, \left[\partial_\mu \omega \right] = 0, \,\,\, \left[\partial_\mu \varphi \right] = 0.
%\end{align}
We can now summarize the junction conditions for the three scalar fields as:
\begin{align}
    \left[\partial_\mu G \right] = 0, \,\,\, \left[\partial_\mu \omega \right] = 0, \,\,\,  \left[\partial_\mu \varphi \right] = 0. \label{ScalarCond}
\end{align}

%%%%%%%%%%%%%%%%%%%%%%%%%%%%%%%%%%%%%%%%%%%%%%%%%%%%%%%%%%%%%%%%%%%%%%%%%%%%%%%%%%%%%%%%%%%%%%%%%%%%%%%%%%%%%%%%%
\subsection{Junction Conditions for the Vector Field Strength Tensor}

Near the junction we can write the vector field as
\begin{align}
    A_\mu = \Theta (\ell) \, A_\mu^+ + \Theta (-\ell) \, A^-_\mu.
\end{align}
Due to the possible presence of non-linear terms as $A^\mu A_\mu$, in the potential $V_A(A_\mu)$, the above expression does not contain any  $\delta$-part. If we include the $\delta$-part then it will produce ill defined products of distributions in the equation of motion for the $A_\mu$ field. The derivative of the above expression is:
\begin{align}
    \partial_\nu A_\mu = \Theta (\ell) \, \partial_\nu A_\mu^+ + \Theta (-\ell) \, \partial_\nu A^-_\mu + \epsilon \delta (\ell) \left[ A_\mu \right] n_\nu.
\end{align}
Using the basic definition for $B_{\mu \nu}$ and using the above expression, we get
\begin{align}
    B_{\mu \nu} = \Theta (\ell) B_{\mu \nu}^+ + \Theta (-\ell) &B_{\mu \nu}^- \nonumber \\  &+ \epsilon \delta (\ell) \left( \left[A_\mu \right] n_\nu  - \left[A_\nu \right] n_\mu \right).
\end{align}
As in the expression of $\mathscr{F}$, there is a term quadratic in $B_{\mu \nu}$, unless we assume that the $\delta$-part of $B_{\mu \nu}$ is zero, there will be terms involving the product of $\Theta$ and $\delta$ in the equation of motion of $\omega$. Consequently we impose:
\begin{align}
    \left[A_\mu \right] n_\nu - \left[A_\nu \right] n_\mu  = 0.  \label{cond_Bvec}
\end{align}
%The above relation can be satisfied if we assume
%\begin{align}
%     \left[A_\mu \right] = \mathscr{A}_{\perp} n_\mu, \label{cond_vecA}
%\end{align}
%for some $\mathscr{A}_{\perp}$ which is in general a function of the hypersurface coordinates. This means that, the %discontinuity in the vector field along the hypersurface is along  the normal direction. We can get $\mathscr{A}_{\perp}$ by %taking the inner product of the above relation with $n^\nu$:
%\begin{align}
%    \mathscr{A}_{\perp} \equiv n^\mu \left[A_\mu \right].
%\end{align}
%
%The simplest way to satisfy this condition, is to assume
%\begin{align}
%    \left[A_\mu \right] = 0. \label{2cond_Bvec}
%\end{align}
If the above condition holds, then we have
\begin{align}
    B_{\mu \nu} = \Theta (\ell) B_{\mu \nu}^+ + \Theta (-\ell) B_{\mu \nu}^- \label{B_dist}\,,
\end{align}
which yields:
%By applying the law of covariant derivative, we can write
%\begin{align}
%    \nabla_\rho B^{\mu \nu} = \partial_\rho B^{\mu \nu} + \Gamma^\mu_{\rho \lambda} B^{\lambda \nu} + \Gamma^\nu_{\rho \lambda} %B^{\mu \lambda}.
%\end{align}
%Now taking the derivative of Eq.~\eqref{B_dist} and using the form in Eq.~\eqref{christoffel_dist} we have
\begin{align}
    \nabla_\nu B^{\mu \nu} = \Theta (\ell) \, \nabla_\nu B^{\mu \nu}_+ + \Theta (- \ell) \, &\nabla_\nu B^{\mu \nu}_- \nonumber \\ &+ \epsilon \delta (\ell) \left[ B^{\mu \nu} \right] n_\nu.
\end{align}
If we assume that the 4-current $J^\mu$ has a $\delta$-part, i.e.,
\begin{align}
    J^\mu = \Theta (\ell) J^\mu_+ + \Theta (-\ell) J^\mu_- + \delta (\ell) K^\mu,
    \label{jmue}
\end{align}
then, from the equation of motion of the vector field, we equate the $\delta$-part to write
\begin{align}
    \epsilon \left[B^{\mu \nu} \right] n_\nu = - \frac{1}{\omega} K^\mu.
\end{align}
Taking the projection of the above relation on the hypersurface we have
\begin{align}
    \epsilon e^\mu_a \left[{B_\mu}^\nu \right] n_\nu = - \frac{1}{\omega} K_\mu e^\mu_a = - \frac{1}{\omega} K_a. \label{B_junc_cond}
\end{align}
The component $K_1$ has the interpretation of induced surface charge density on the hypersurface and we will denote it by $\mathscr{Q}$.

%We can take the inner product of Eq~\eqref{cond_Bvec} with $e^\nu_a$ and write
%\begin{align}
%    \left[A_\mu \right] n_\nu e^\nu_a = \left[A_\nu \right] e^\nu_a n_\mu \implies e^\mu_a \left[A_\mu \right] = 0. \label{2cond_Bvec}
%\end{align}
%This relation is just another way of saying that the discontinuity in $A_\mu$ is solely along the normal direction to the hypersurface $\Sigma$.

%%%%%%%%%%%%%%%%%%%%%%%%%%%%%%%%%%%%%%%%%%%%%%%%%%%%%%%%%%%%%%%%%%%%%%%%%%%%%%%%%%%%%%%%%%%%%%%%%%%%%%%%%%%%%%
\section{The interior and exterior spacetimes involved in the gravitational collapse process}
\label{sptimes}

In the gravitational collapse process, presented in this paper, the interior spacetime is assumed to be described by the FLRW-like solution in STVG theory and the external spacetime is assumed to be described by the static and spherically symmetric RN-like solution in STVG theory. As the present theory contains many kinds of fields it is very difficult to model the exterior spacetime as a Schwarzschild spacetime, it will be difficult to accommodate the other fields in it. On the other hand, if the scalar fields are suitably chosen to be constant or non-dynamical, only keeping the zeroth component of the 4-vector field $A_0(r)$ in the RN-like spacetime one can model the external black hole solution. The interior FLRW can in principle accommodate the zeroth component of the 4-vector field $A_0(\tau)$. It is seen that the junction condition for the zeroth component of this 4-vector becomes ill defined on the hypersurface $\Sigma$ and hence we will not match the zeroth component of the 4-vector field on the junction, it in general can change discontinuously through the junction. To keep things simple, we will assume that in the interior FLRW-like spacetime all the scalar fields are non-dynamical, and even the $A_0(\tau)$ component is assumed to be zero. In such a case the $A_0(r)$ field in the RN-like spacetime originates from a STVG-charged shell on $\Sigma$ which carries a charge density as given in Eq.~(\ref{B_junc_cond}).  The RN-like spacetime is assumed to be devoid of matter and consequently the STVG current, produced by matter and the 4-vector coupling, $J^\mu=0$ outside. Inside the FLRW spacetime we will assume baryonic matter and dark-energy-like component but here $A_0(\tau)$ is assumed to be zero and hence in the interior spacetime also we have $J^\mu=0$. As a consequence in the proposed collapse model the STVG-current can only reside on the shell separating the two spacetimes and this current comes from the term proportional to the $\delta(\ell)$ term in the expression of $J^\mu$ in Eq.~(\ref{jmue}). On the shell the energy-momentum tensor $S_{\mu \nu}$ is produced by interaction of matter and $A_0$ in such a way that $S_{\mu \nu}$ remains traceless. In this article we will not explicitly write the interaction Lagrangian of the matter component and $A_0$, we will only specify this interaction with some prescribed form of $S_{\mu \nu}$. 

In a simple and minimalist model, where the internal spacetime is practically the FLRW spacetime and the external spacetime is the RN-like spacetime in STVG theory one may ignore the dynamics of the gravitational coupling, $G$, and assume it to be constant throughout the spacetime manifold. This constant nature of $G$ can be interpreted in multiple ways and we will specify these ways in the next section.
%%%%%%%%%%%%%%%%%%%%%%%%%%%%%%%%%%%%%%%%%%%%%%%%%%%%%%%%%%%%%%%%
\subsection{Exterior Static and Spherically Symmetric Spacetime}

First, we will discuss the exterior spacetime solution of the field equations in a vacuum. For simplicity, we take the cosmological constant $\Lambda$ to be zero, i.e.,
\begin{align}
    T^{(M)}_{\mu \nu} = 0, \,\,\, \text{and} \,\,\, \Lambda = 0.
\end{align}
A generic static and spherically symmetric spacetime has a metric of the following form:
\begin{align}
      \text{d}s_+^2 = - e^{2 \alpha (r)} \text{d}t^2 + e^{2 \beta (r)} \text{d}r^2 + r^2 \text{d} \Omega^2, \nonumber \\
      \text{where}, \,\, \text{d} \Omega^2 \equiv \left( \text{d} \vartheta^2+\sin^2 \vartheta \, \text{d} \phi^2 \right). \label{SphSymmMet}
\end{align}
In the exterior region we take all the scalar fields as constants and only $A_\mu(r)$ is the nontrivial 4-vector field present there. For a proper RN-like solution outside we assume vanishing potentials for the scalar fields and the vector field, i.e., $V_X$ $(X=G,\,\omega,\,\varphi,\,A_\mu)$. In such a case the field equation Eq.~\eqref{FE} reads: 
\begin{align}
    R_{\mu \nu} = 8 \pi G \prescript{(V)}{}{T}_{\mu \nu},
\end{align}
as the trace of $\prescript{(V)}{}{T}_{\mu \nu}$ vanishes and consequently the Ricci scalar also vanishes.

We can calculate the nonzero mixed Ricci components as
\begin{align}
    &R^0_0 = - e^{ - 2 \beta} \left( \alpha^{\prime \prime} + {\alpha^\prime}^2 - \alpha^\prime \beta^\prime + \frac{2 \alpha^\prime}{r} \right), \\
    &R^1_1 = - e^{- 2 \beta} \left( \alpha^{\prime \prime} + {\alpha^\prime}^2 - \alpha^\prime \beta^\prime - \frac{2 \beta^\prime}{r} \right), \\
    &R^2_2 =  R^3_3 = \frac{e^{- 2 \beta}}{r^2} \left\{  r \left(\beta^\prime - \alpha^\prime \right) -1\right\} + \frac{1}{r^2}.
\end{align}
In the above and the following, the labels $0, 1, 2, 3$ stand for $t, r, \vartheta$ and $\phi$ respectively and the prime represents a derivative with respect to the radial coordinate, $r$.
%it can be checked that, the expressions for the non-vanishing components of the Einstein tensor are:
%\begin{align}
%    &G^0_0 = e^{-2\beta} \left(\frac{1}{r^2} - \frac{2 \beta^{\prime}}{r} \right) - \frac{1}{r^2},  \nonumber \\
%    &G^1_1 = e^{-2\beta} \left(\frac{1}{r^2} + \frac{2 \alpha^{\prime}}{r} \right) - \frac{1}{r^2}, \nonumber \\
%    &G^2_2 = G^3_3 = e^{-2\beta} \left(\alpha^{\prime \prime} + {\alpha^{\prime}}^2 - \alpha^{\prime} \beta^{\prime}  + \frac{\alpha^{\prime} -\beta^{\prime}}{r} \right). \label{ETgen}
%\end{align}

As we are considering a static and spherically symmetric spacetime, we assume that the vector field $A_\mu$ depends only on the radial coordinate, i.e., $$A_\mu \equiv A_\mu (r).$$
We will assume that, the only nonzero component is $A_0$ as the system can only accommodate an STVG-charge which is analogous to the electric charge in the standard RN theory in GR. In the absence of any magnetic field like configuration the system can only have a finite  $A_0(r)$. From Eq.~\eqref{Bdef} we write the nonzero component of the $B_{\mu \nu}$ as follows:
\begin{align}
    B_{01} = - A_0^\prime.
\end{align}
Using this, we write the contravariant component as 
\begin{align}
    B^{01} = e^{-2\alpha - 2\beta} A^\prime_0\,,
\end{align}
which yields
\begin{align}
    B_{\mu \nu} B^{\mu \nu} = -2 e^{- 2 \alpha - 2 \beta} {A^\prime_0}^2.
\end{align}
The equation of motion for the vector field $\nabla_\nu B^{\mu \nu} = 0$ becomes:
\begin{align}
    \partial_1 \left(r^2 e^{\alpha + \beta} B^{01} \right) = 0.
\end{align}
This relation can be integrated to get
\begin{align}
    A^\prime_0 = e^{\alpha + \beta} \frac{Q}{r^2}, \label{dA+}
\end{align}
where $Q$ is an integration constant which will be equated to the total surface charge on the shell. 

In the present case the non-vanishing components of energy-momentum tensor become
\begin{align}
    \prescript{(V)}{}{T}^0_0 = \prescript{(V)}{}{T}^1_1 = - \prescript{(V)}{}{T}^2_2 = - \prescript{(V)}{}{T}^3_3  = - \frac{\omega Q^2}{2 r^4}. 
    %- \frac{1}{2}  \omega e^{- 2 \alpha - 2 \beta} {A^\prime_0}^2.
\end{align}
Equating the $0$-$0$ and $1$-$1$ components of the field equation we get $\alpha^\prime = - \beta^\prime$. We take $\alpha = - \beta$ setting the integration constant to zero. Substituting this solution in the remaining field equation we get the following differential equation:
\begin{align}
    e^{2 \alpha} \left(\frac{2 \alpha^\prime}{r} + \frac{1}{r^2} \right) - \frac{1}{r^2} = - \frac{4 \pi \omega G Q^2 }{r^4}.
\end{align}
The above equation can be integrated to get the result obtained in Ref.~\cite{MoffatMOG}:
\begin{align}
    e^{ 2 \alpha} = 1  - \frac{2 G M}{r} + \frac{G Q^2}{r^2}, \label{RNmetric}
\end{align}
where $2GM$ is an integration constant, and we have set the value of constant $\omega$ as $(4\pi)^{-1}$. In the RN-like spacetime we assume $G$ to be constant and $\varphi=0$. As specified in Ref.~\cite{MoffatMOG}, the above solution can produce a black hole solution in STVG theory, which is not the simple Schwarzschild black hole as we have explicit presence of the STVG charge here. It is like an RN black hole but one must remember that the field $A_0(r)$ does not correspond to any real electric field, it is a gravitational degree of freedom in STVG theory. 

%%%%%%%%%%%%%%%%%%%%%%%%%%%%%%%%%%%%%%%%%%%%%%%%%%%%%%%%%%%%%%%%%%%%%%%%%%%%%%%%%%%%%%%%%%%%%%%%%%%%%%%%%%%%%%%%%%%%%%
\subsection{Interior FLRW Spacetime}

The interior spacetime is taken to be a closed FLRW spacetime, with the line element
\begin{align}
    \text{d}s_-^2
    = - \text{d}\tau^2
    + a^2 \left(\tau\right)
    \left(\text{d}\chi^2 + \sin^2 \chi \, \text{d} \Omega^2 \right).
    \label{FLRWmet}
\end{align}

In this section, a dot or prime over a quantity represents a derivative with respect to $\tau$ or $\chi$ respectively. As this is a homogeneous spacetime, every physical quantity (e.g., fields) will depend only on the temporal coordinate $\tau$. 

As we are considering a homogeneous spacetime, we assume that the vector field $A_\mu$ depends only on the temporal coordinate, i.e., $A_\mu \equiv A_\mu (\tau)$. Isotropy requires:
\begin{align}
   A^0 \equiv A^0 (\tau), \,\,\, A^i = 0.
\end{align}
As the tensor $B^{\mu \nu}$ vanishes in $\mathscr{V}^-$, Eq.~\eqref{EoM_A} implies 
\begin{align}
    J^\mu = 0, \label{charge_def_int}
\end{align}
and consequently the conservation law for $J^\mu$ (Eq.~\eqref{divJZero}) and Eq.~\eqref{divEoMA} are automatically satisfied as $V_A(A^0)$ is assumed to be zero. In the present case we are assuming that $\omega$ and $G$ are constants in the interior region whereas $\varphi=0$ inside. In the simplest model of collapse we also assume $A^0(\tau)=0$ in the interior region. 

In the interior spacetime we assume there is a dark-energy like component with the energy-momentum tensor:
\begin{align}
    \prescript{\left(DE\right)}{}T_{\mu \nu} = \left(\rho_{DE} + p_{DE} \right) U_\mu U_\nu + p_{DE} g_{\mu \nu}\,.
\end{align}
In addition to this, we will assume that there is an ideal fluid in the FLRW spacetime, whose stress tensor is
\begin{align}
    \prescript{\left(M\right)}{}T_{\mu \nu} = \left(\rho_M + p_M \right) U_\mu U_\nu + p_M g_{\mu \nu},
\end{align}
where $U^\mu \equiv \left(1, 0, 0, 0 \right) $ is the 4-velocity of the comoving fluid in the comoving coordinates we are working in (Eq.~\eqref{FLRWmet}). The components of $T^{\left(M\right)}_{\mu \nu}$ are
\begin{align}
    \prescript{\left(M\right)}{}T^0_0 = - \rho_M, \,\,\, \prescript{\left(M\right)}{}T^1_1 = \prescript{\left(M\right)}{}T^2_2 = \prescript{\left(M\right)}{}T^3_3 = p_M, 
\end{align}
and similar expressions for $\prescript{\left(DE\right)}{}T^\mu_\nu$.
We can now write the field equations. The 0-0 component reads
\begin{align}
     3 \left(\frac{\dot{a}^2}{a^2} + \frac{1}{a^2} \right)  = 8 \pi G \rho. \label{FridE1}
\end{align}
Similarly the 1-1, 2-2 and 3-3 components yield
\begin{align}
     -\left( \frac{2 \ddot{a}}{a}\right. + &\left.\frac{\dot{a}^2}{a^2} + \frac{1}{a^2} \right) = 8 \pi G  p. \label{FridE2}
\end{align}
In the above two equations, we have defined the total energy density and total pressure as $\rho = \rho_M + \rho_{DE} $ and $p = p_M + p_{DE} $. In the interior FLRW spacetime we have also assumed that all the potentials $V_X$ $(X=G,\,\omega,\,\varphi,\,A_\mu)$
vanish, just as in the external spacetime.

%%%%%%%%%%%%%%%%%%%%%%%%%%%%%%%%%%%%%%%%%%%%%%%%%%%%%%%%%%%%%%%%%%%%%%%%%%%%%%%%%%%%%%%%%%%%%%%
\subsection{Imposing the Junction Conditions}

The metric when looked from the outside ($\mathscr{V}^+$) has the form given by Eq.~\eqref{SphSymmMet} with $e^{2\alpha (r)} = e^{-2 \beta (r)} = f (r) $, and written as  
\begin{align}
      &\text{d}s_+^2 = - f \text{d}t^2 + f^{-1}  \text{d}r^2 + r^2  \text{d} \Omega^2, \nonumber \\
      &\text{with}, \,\, f(r) \equiv e^{2 \alpha (r)} = \left(1 - \frac{2 G M}{r} + \frac{G Q^2}{r^2} \right). \label{SchwarzMet}
\end{align}
As seen from the outside, the parametric equation of the hypersurface is $r = R (\tau)$\footnote{There can be confusion at this point as $R$ here is not the Ricci scalar, also represented by the same symbol. As henceforth the Ricci scalar never appears explicitly we will use the same symbol for the physical radial distance to the shell in this article.}, $t = T (\tau)$. Here $\tau$ is the same proper time-coordinate appearing in Eq.~\eqref{FLRWmet}. 

As the coordinate on $\Sigma$, we choose $y^a \equiv (\tau, \vartheta, \phi)$. We can now calculate the induced metric. As seen from $\mathscr{V^-}$, the induced metric can be written as:
\begin{align}
    \text{d}s_{\Sigma}^2 = - \text{d}\tau^2 + a^2 (\tau) \sin^2 \chi_0 \,  \text{d} \Omega^2. 
\end{align}
In the above expression, $\chi_0$ is the value of $\chi$ on the hypersurface $\Sigma$. As seen from the region $\mathscr{V}^+$, the induced metric is
\begin{align}
    \text{d} s^2_\Sigma = - \left( F \dot{T}^2  - F^{-1} \dot{R}^2 \right) \text{d}\tau^2 + R^2 \text{d}\Omega^2,
\end{align}
where we have denoted $F \equiv 1 - 2 G M / R + GQ^2/R^2$, which is the value of $f$ on $\Sigma$. From the first junction condition, the induced metric on both sides of the hypersurface is same, hence:
\begin{align}
    R (\tau) = a (\tau) \sin \chi_0, \,\,\, F \dot{T}^2 - F^{-1} \dot{R}^2 = 1. \label{1stJunConds}
\end{align}
The second condition implies
\begin{align}
    F \dot{T} = + \sqrt{F + {\dot{R}}^2} \equiv \beta (R, \dot{R}). \label{betadef}
\end{align}
We can calculate the unit normal $n^\mu$ on $\Sigma$ from the following two relations: $u^\mu n_\mu = 0$ and $n^\mu n_\mu = + 1$. Here $u^\mu$ ($ \equiv dx^\mu / d \tau$) is the four velocity of a dust particle  on the boundary surface which always stays on the hypersurface $\Sigma$. As seen from $\mathscr{V}^-$, we have $u^\mu_- \equiv (1, 0, 0, 0)$ and $n^-_\mu \equiv (0, a, 0, 0)$. Now, for an observer in $\mathscr{V}^+$, we have $u^\mu_+ \equiv (\dot{T}, \dot{R}, 0, 0)$ and the unit normal can be calculated by the aforementioned two relations as  $n^+_\alpha \equiv (- \dot{R}, \dot{T}, 0, 0)$.

To implement the condition on $B_{\mu \nu}$, as coming from Eq.~\eqref{B_junc_cond}, we first note that $\epsilon = n^\mu n_\mu=1$ specifying that $\Sigma$ is a timelike hypersurface.  In this case the junction condition (for index $a=1$) is: 
\begin{align}
    e^0_1 {B^{+}_0}^1 n^+_1 + e^1_1 {B^{+}_1}^0 n^+_0 - e^0_1 {B^{-}_0}^1 n^-_1 =  - \frac{1}{\omega} \mathscr{Q}.
\end{align}
The first two terms on the LHS are $(F \dot{T}^2 - F^{-1} \dot{R}^2) B^+_{01}$, which equal $B^+_{01}$, using the equality of the first fundamental form. The third term can be evaluated to be $(B^{-}_{01}/\sin \chi_0)$. Using these inputs the junction condition becomes
\begin{align}
    B^{+}_{01} - (\sin\chi_0)^{-1} B^{-}_{01} = - \frac{1}{\omega} \mathscr{Q}. \label{B_junc_cond2}
\end{align}
Using the solution for $A^\prime_0$ in $\mathscr{V}^+$  obtained from Eq.~\eqref{dA+} and the above relation we get: 
\begin{align}
    \mathscr{Q} = \frac{Q}{4 \pi {R}^2(\tau)}.   
\end{align}
This relates the STVG-charge parameter $Q$ with the induced surface charge density $\mathscr{Q}$.

Next, we evaluate the conditions imposed on the scalar fields given in Eq~\eqref{ScalarCond}. To evaluate these conditions, we introduce the coordinates $\{x^\mu\}$, which represents an overlapping coordinate system, which is used in the neighborhood of $\Sigma$ and overlaps with $\{x^\mu_+\}$ and $\{x^\mu_-\}$ in $\mathscr{V}^+$ and $\mathscr{V}^-$, respectively. In particular, we choose $x^\mu \equiv \{\iota, \zeta, \vartheta, \phi\}$ for the evaluation of the junction conditions. Here $\iota$ is a timelike coordinate and the remaining coordinates are spacelike . For the temporal component, we have 
\begin{align}
 {\partial }_0 G_+ &={\partial}_\iota G_+  = \left(\frac{\partial x^\mu_+}{\partial {x}^0} \right) \partial_{\mu} G_+  = \left(\frac{\partial r }{\partial \iota} \right)  G_+^\prime, \nonumber \\
{\partial}_0 G_- &=   {\partial }_\iota G_- = \left(\frac{\partial x^\mu_-}{\partial {x}^0} \right) \partial_{\mu} G_-  = \left(\frac{\partial \tau }{\partial \iota} \right) \dot{G}_-.
\end{align}
So, the condition that $\left[{\partial}_0 G \right]$ vanishes, implies
\begin{align}
    \dot{R} G_+^\prime = \dot{G}_-. \label{2Gcond}
\end{align}
This relation determines the time variation of the scalar field $G (\tau)$ in the interior FLRW spacetime given the expression for $R (\tau)$ and $G (r)$ in $\mathscr{V}^+$.  Similarly
\begin{align}
 {\partial}_1 G_+ &=   {\partial }_\zeta G_+ = \left(\frac{\partial x^\mu_+}{\partial {x}^1} \right) \partial_{\mu}G_+ = \left(\frac{\partial r }{\partial \zeta} \right)  G_+^\prime, \nonumber \\
 {\partial}_1 G_- &=   {\partial }_\zeta G_- = \left(\frac{\partial x^\mu_-}{\partial {x}^1} \right) \partial_{\mu}G_- = \left(\frac{\partial \tau }{\partial \zeta} \right)  \dot{G}_-.
\end{align}
Hence, $\left[{\partial}_1 G \right] = 0$ also yields the same condition as in Eq.~\eqref{2Gcond} and it can be checked that the other components do not produce any relation. For the scalar fields $\omega$ and $\varphi$, we get the same condition as in Eq.~\eqref{2Gcond}. 

One can see that, the above relations can also be obtained from the continuity of the scalar fields, viz.,
\begin{align}
    G_- (\tau) = G_+ (R (\tau)) \implies \dot{G}_- (\tau) &= \frac{\partial G_+ (R)}{\partial R} \frac{d R (\tau)}{d \tau} \nonumber \\  
    &= \dot{R} G^\prime_+,
\end{align}
and similarly for the other fields. So we summarize the junction conditions for the scalar fields once again:
\begin{align}
    \dot{G}_- = \dot{R} \, G^\prime_+, \,\,\,\dot{\omega}_- = \dot{R} \, \omega^\prime_+, \,\,\,\dot{\varphi}_- = \dot{R} \, \varphi^\prime_+. \label{3ScalarCond}
\end{align}
These conditions allow $G,\omega,\varphi$ to be constant on both sides of the junction.

%%%%%%%%%%%%%%%%%%%%%%%%%%%%%%%%%%%%%%%%%%%%%%%%%%%%%%%%%%%%%%%%%%%%%%%%%%%%%%%%%%%%%%%%%%%%%%%%%%%%%%%%%%%%%%%%%%%
\section{Solutions with Thin Shells} 
\label{NumSol}

In this article we present the simplest workable model of a gravitational collapse which can be a generalization of the Oppenheimer-Snyder collapse in STVG theory. Due to the presence of so many fields in this model of gravity we have to judiciously keep some of them non-dynamical so that the system remains manageable. In this scheme we have practically frozen the dynamics of all the scalar fields and we have also set the vector field to be zero inside the internal FLRW spacetime. The external RN-like spacetime has a nonzero $A_0(r)$ which is produced from the distribution of STVG-charge on the shell at  the junction between the two spacetimes. The shell is gravitationally charged and it has its own energy-momentum tensor which is traceless.  

In the present case there are two possible ways in which the collapse can be conceived. These two ways can produce different forms of collapse. The first possibility relies on the non-dynamic nature of the effective gravitational coupling, $G$. In this case one can go one step further and disregard the equation of motion of the effective $G$ field and its matching conditions, consequently Eq.~(\ref{Gcond}) can also be disregarded. Strictly speaking this is not the full STVG theory but a simplified truncated version of it. In this truncated picture, $G$ attains an effective constant value, as discussed in Ref.~\cite{MoffatMOG}, and does not have the status of a scalar field. In STVG theory the effective constant value of the coupling $G$ arises from observational requirements \cite{Moffat:2013sja}. In such a case the shell separating the two spacetimes can have an energy-momentum tensor which is not traceless, i.e., $S\ne 0$. In the second possibility, one keeps the possible field status of $G$ but assumes $G$ to be effectively a constant throughout the spacetime with an effective value slightly differing from the universal gravitational constant as discussed in Ref.~\cite{MoffatMOG}. In this case one does not disregard the field equations of $G$ or its matching condition and hence Eq.~(\ref{Gcond}) cannot be disregarded. One can check that an effectively constant $G$ does not contradict any of the junction conditions. As a consequence, in this second possibility of collapse, one must have $S=0$. Which of these two possibilities is realized in nature can only be decided observationally.

We present collapsing solutions in both the scenarios. It is seen that in the truncated STVG theory if $S\ne 0$ one does have freedom to choose the equation of state for the interacting matter on the shell and this scenario produces a subextremal RN-like black hole in STVG theory. On the other hand if we assume $S=0$ during collapse, one may produce an extremal RN-like black hole in STVG theory. 

%%%%%%%%%%%%%%%%%%%%%%%%%%%%%%%%%%%%%%%%%%%%%%%%%%%%%%%%%%%%%%%%%%%%%%%%%%%%%%%%%%%%%%%%%%%%%%%%%%%%%%%%%%%%%%%%%%%%%%%
\subsection{Interior solution with Dark Energy and a Dust-like Matter Component, $S\ne 0$ }
%%%%%%%%%%%%%%%%%%%%%%%%%%%%%%%%%%%%%%%%%%%%%%%%%%%%%%%%%%%%%%%%%%%%%%%%%%%%%%%%%%%%%%%%%%%%%%%%%%%%%%%%%%%%%%%%%%%%%%%%%%%%%%%%%%%%%%%%%%%%%%%%%%%%%%%%%%

\begin{figure*}
\begin{subfigure}{.5\textwidth}
  \centering
  \includegraphics[width=1.0\linewidth]
  {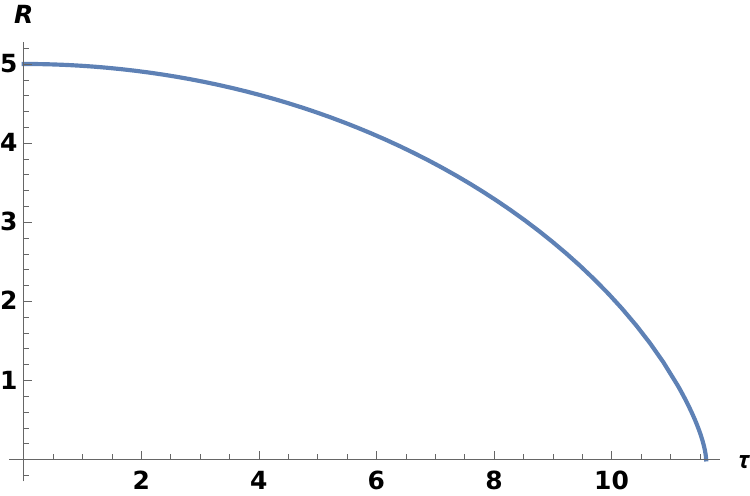}
  \caption{}
  \label{ScaleFactor}
\end{subfigure}%
\begin{subfigure}{.5\textwidth}
  \centering
  \includegraphics[width=1.0\linewidth]{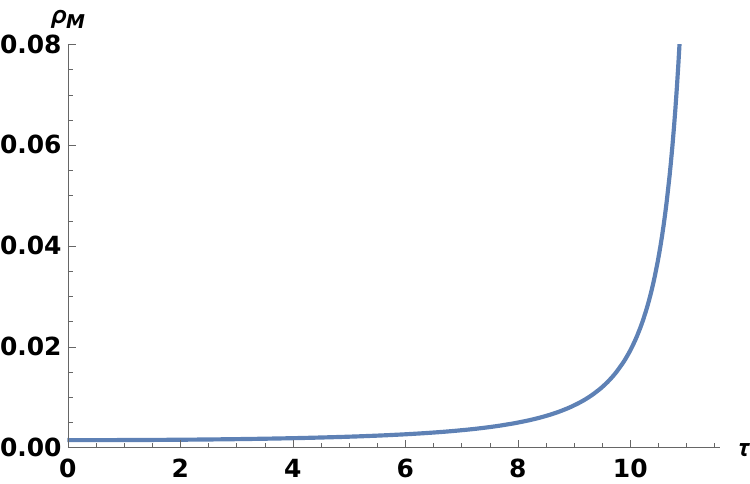}
  \caption{}
  \label{shell_energy_density}
\end{subfigure}%
\newline
\begin{subfigure}{.5\textwidth}
  \centering
  \includegraphics[width=1.0\linewidth]{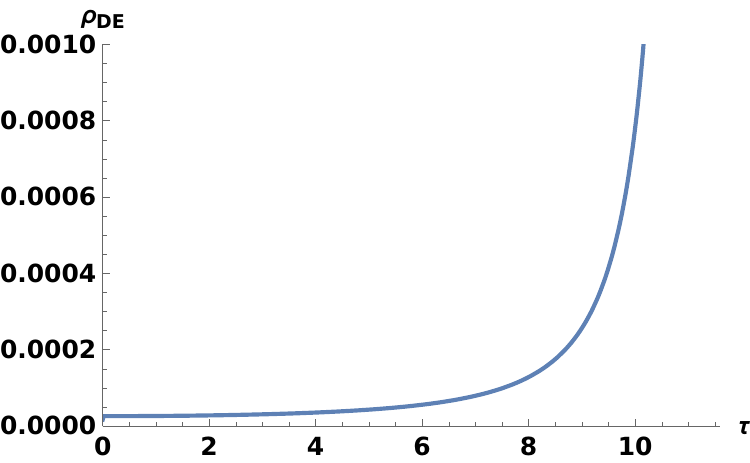}
  \caption{}
  \label{rhoM}
\end{subfigure}%
\begin{subfigure}{.5\textwidth}
  \centering
  \includegraphics[width=1.0\linewidth]{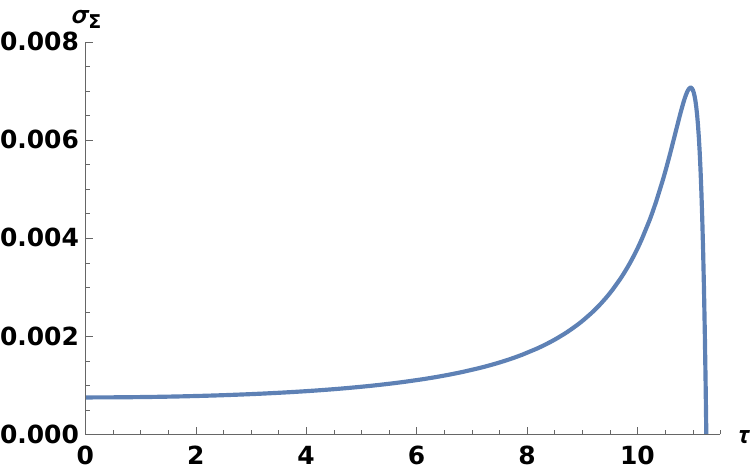}
  \caption{}
  \label{ShellEnergyDensity}
\end{subfigure}%
\caption{The above panel shows the variation of $R (\tau) $ (Fig.~[\ref{ScaleFactor}]), matter-energy density $\rho_M$ (Fig.~[\ref{shell_energy_density}]), energy density $\rho_{DE}$ (Fig.~[\ref{rhoM}]) and the energy density of the shell $\sigma_\Sigma$ with proper time (Fig.~[\ref{ShellEnergyDensity}]) in the interior spacetime region.  Here we have taken the  values of various constants and parameters as follows: $\sin\chi_0 = 1/\sqrt{2}$, $G_0 = 1$, $M = 1$, $\alpha = 0.6$ and $Q^2 = \alpha G_0 M^2 $ and $G = G_0 (1 + \alpha)$.  The initial conditions are chosen such that, at $\tau = 0$, we have $R = 5$ and $\dot{R} = 0$. 
} 
\label{gen_Dust_IntSol}
\end{figure*}
%%%%%%%%%%%%%%%%%%%%%%%%%%%%%%%%%%
%%%%%%%%%%%%%%%%%%%%%%%%%%%%%%%%%%%%%%%%%%%%%%%%%%%%%%%%%%%%%
\begin{figure*}
 \centering
\includegraphics[width=0.5\linewidth]{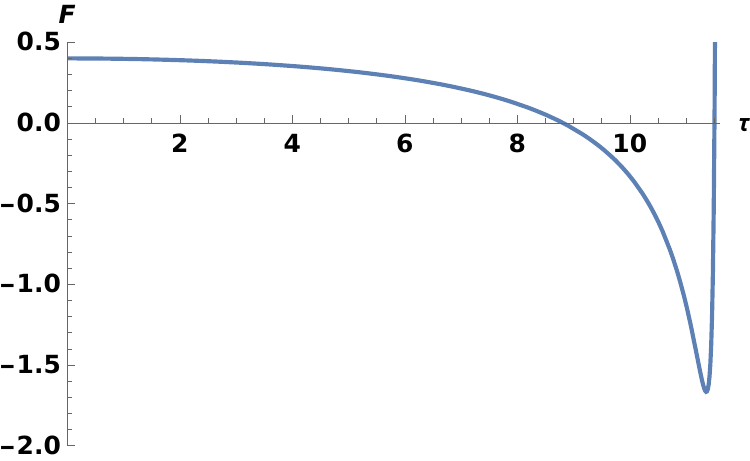}
\caption{The above figure shows the variation of the horizon function, $F$, with respect to the proper time.
}
\label{HorizonFunction}
\end{figure*}%\cite{Famaey:2011kh}
%%%%%%%%%%%%%%%%%%%%%%%%%%%%%%%%%%%%%%%%%%%%%%%%%%%%%%%%%%%%%%%%%%%%%%%%%%%%%%%%%%%%%%%%%%%%%%%%%%

In the first case, we disregard the field status of $G$ and focus on a truncated version of the original STVG theory we proposed initially. To find out how the gravitational collapse commences, we first solve the dynamics inside the FLRW spacetime core. We will solve the Friedmann equations (Eqs.~\eqref{FridE1} and ~\eqref{FridE2}) in $\mathscr{V}^-$ assuming the matter component as a pressureless dust, i.e., $p_M = 0$.  The interior patch can also have dark energy component with energy density $\rho_{DE}$ and pressure $p_{DE}$. The relevant equations are:
\begin{align}
    &\rho_{DE} + \rho_M =  \frac{3}{8 \pi G } \left(\frac{\dot{a}^2}{a^2} + \frac{1}{a^2} \right), \label{RhoM}  \\
    &p_{DE} = - \frac{1}{8 \pi G}\left( \frac{2 \ddot{a}}{a} + \frac{\dot{a}^2}{a^2} + \frac{1}{a^2} \right). \label{pM} 
\end{align}
We further impose that for DE component, we have $p_{DE} = - \rho_{DE}$.
%%%%%%%%%%%%%%%%%%%%%%%%%%%%%%%%%%%%%%%%%%%%%%%%%%%%%%%%%%%%%%%%%%%%%%%%%%%%%%%%%

%From Eq.~\eqref{FridE1} and ~\eqref{FridE2} we can calculate the energy density and pressure in $\mathscr{V}^-$ as follows:
%\begin{align}
%    \rho_M &=  \frac{3}{8 \pi G } \left(\frac{\dot{a}^2}{a^2} + \frac{1}{a^2} \right) - \widetilde{V} , \label{RhoM}  \\
%    p_M &= - \frac{1}{8 \pi G}\left( \frac{2 \ddot{a}}{a} + \frac{\dot{a}^2}{a^2} + \frac{1}{a^2} \right) + \widetilde{V}. \label{pM} 
%\end{align}
%In the above we have defined
%\begin{align}
%    \widetilde{V} \equiv \frac{1}{G} \left( \frac{V_G}{G^2} + V_\omega + \frac{V_\varphi}{\varphi^2}\right), 
%\end{align}
%which is a constant due to the constancy of the scalar fields $G$, $\omega$ and $\varphi$.
%%%%%%%%%%%%%%%%%%%%%%%%%%%%%%%%%%%%%%%%%%%%%%%%%%%%%%%%%%%%%%%%%%%%%%%%%%%%%%%%%%%%%%%%%%%%%%%%%%%%%%%%%%

We take the stress-energy tensor corresponding to the shell as
\begin{align}
    S^a_b = {\rm{diag}}(-\sigma_\Sigma,\, p_\Sigma,\, p_\Sigma).
\end{align}
In the above, $\sigma_\Sigma$ and $p_{\Sigma}$ are the energy density and pressure corresponding to the shell where $S^\tau_\tau = - \sigma_\Sigma$ and $S^\vartheta_\vartheta = p_\Sigma = S^\phi_\phi$ and $S = 2 p_\Sigma - \sigma_\Sigma \ne 0$. 

To impose the second junction condition, we rewrite Eq.~\eqref{2ndJC} as
\begin{align}
    \left[K_{ab} \right] = - 8 \pi G \left(S_{ab} - \frac{1}{2} S h_{ab}  \right)\,.\label{JCShell}
\end{align}
Calculating the extrinsic curvature from both the sides of the shell, we have:
\begin{align}
    &K^\tau_{+ \tau} = \frac{\dot{\beta}}{\dot{R}}, \,\,\, K^\vartheta_{+ \vartheta} = K^\phi_{+ \phi} = \frac{\beta}{R}; \nonumber \\
     &K^\tau_{- \tau} = 0, \,\,\, K^\vartheta_{- \vartheta} = K^\phi_{- \phi} = \frac{\cot\chi_0}{a} = \frac{\cos\chi_0}{R}.
\end{align}
Using these values we obtain the two following equations determining the dynamics of the shell:
\begin{align}
    \frac{\dot{\beta}}{\dot{R}} = 8 \pi G \left(\frac{1}{2} \sigma_\Sigma + p_{\Sigma} \right),  \label{shellPressure} \\
    \frac{\beta}{R} - \frac{\cos\chi_0}{R} = -4 \pi G \sigma_\Sigma, \label{shellEnergy}
\end{align}
which can be combined to write
\begin{align}
     \frac{\dot{\beta}}{\dot{R}} +
    \frac{\beta}{R} - \frac{\cos\chi_0}{R} = 8 \pi G p_\Sigma\,. \label{DiffEqR}
\end{align}
This  is a second-order differential equation for $R$ which is a function of $\tau$. From Eq.~\eqref{betadef}, we can see that, $\beta$ depends on $M$ through $F$, and consequently the initial value of the total energy density in the interior region is fixed as the values of the mass parameter $M$ of the external RN spacetime is fixed. The subsequent development of the total energy density and the scale factor are determined by the above equations.

To solve the dynamics of collapse we  have five unknowns, $R(\tau) $, $\rho_M$,  $\rho_{DE}$, $\sigma_\Sigma$ and $p_\Sigma$ and four equations (Eqs.~\eqref{RhoM}, \eqref{pM}, \eqref{shellPressure}, \eqref{shellEnergy}). Consequently, we have to assume some functional form for the pressure or energy density of the shell. There can be various possibilities, which arise from various kinds of interaction of matter and the $A_0$ field on the shell. From a purely phenomenological point of view, we assume that $p_\Sigma$ is equal to $p_{DE}$. This choice is not unique and one may have chosen a different form for shell pressure. In this simplistic model we proceed with this choice. Now we can solve Eq.~\eqref{DiffEqR}  to get the solution for $R(\tau)$, which we can use to calculate shell-energy density and matter energy density from Eqs.~\eqref{shellEnergy} and \eqref{RhoM}. The results are shown in Fig.~\ref{gen_Dust_IntSol}.

\begin{figure*}
\begin{subfigure}{.5\textwidth}
  \centering
  \includegraphics[width=1.0\linewidth]
  {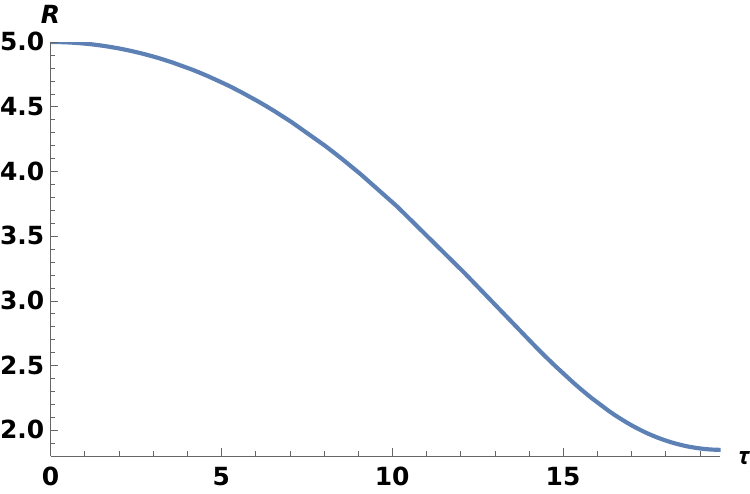}
  \caption{}
  \label{ScaleFactorm}
\end{subfigure}%
\begin{subfigure}{.5\textwidth}
  \centering
  \includegraphics[width=1.0\linewidth]{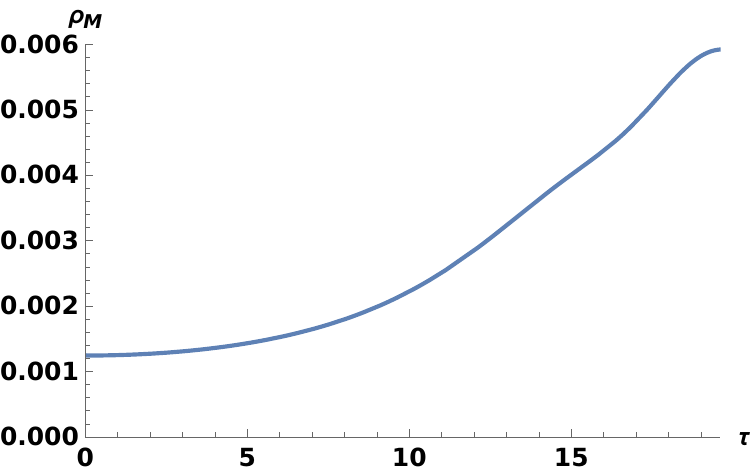}
  \caption{}
  \label{shell_energy_densitym}
\end{subfigure}%
\newline
\begin{subfigure}{.5\textwidth}
  \centering
  \includegraphics[width=1.0\linewidth]{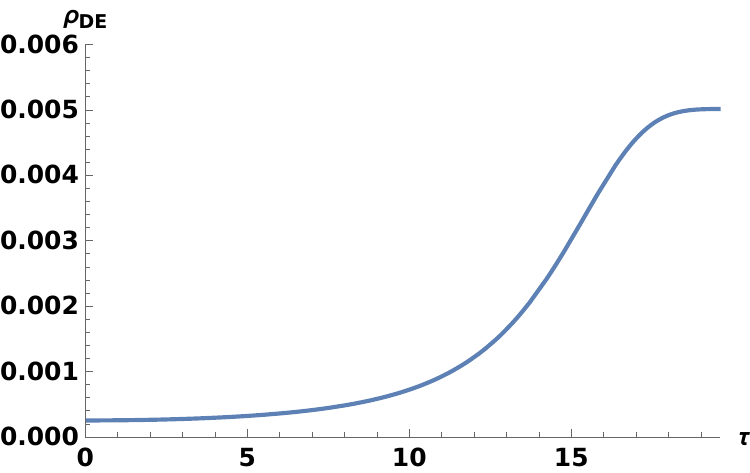}
  \caption{}
  \label{rhoMm}
\end{subfigure}%
\begin{subfigure}{.5\textwidth}
  \centering
  \includegraphics[width=1.0\linewidth]{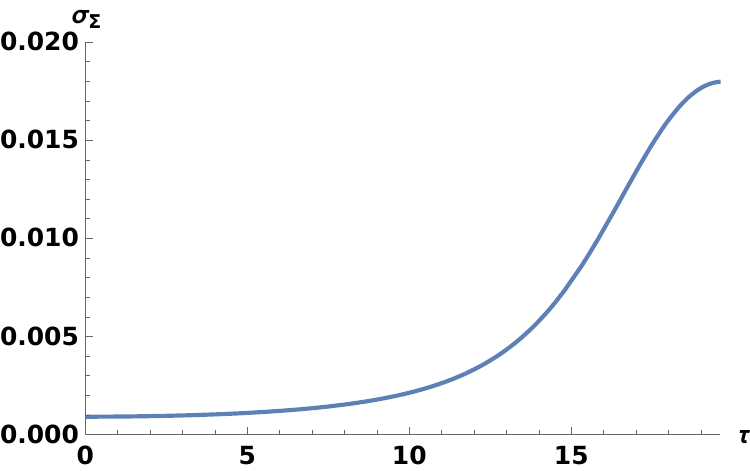}
  \caption{}
  \label{ShellEnergyDensitym}
\end{subfigure}%
\caption{The above panel shows the variation of $R (\tau) $ (Fig.~[\ref{ScaleFactorm}]), matter-energy density $\rho_M$ (Fig.~[\ref{shell_energy_densitym}]), energy density $\rho_{DE}$ (Fig.~[\ref{rhoMm}])  and the energy density of the shell $\sigma_\Sigma$ with proper time (Fig.~[\ref{ShellEnergyDensitym}]) in the interior spacetime region.  Here we have taken the  values of various constants and parameters as follows: $\sin\chi_0 = 1/\sqrt{2}$, $G_0 = 1$, $M = 1.2$, $G = 1.6$ and $Q^2 = G M^2$. This corresponds to an extremal RN-like solution in the exterior spacetime region. For these values of parameters, we have $Q = 2.304$ and the initial conditions are chosen such that, at $\tau = 0$, we have  $R = 5$ and $\dot{R} = 0$. This solution corresponds to the case, when the trace of the stress-energy tensor of the shell vanishes which implies $p_\Sigma = (1/2)\sigma_\Sigma$. 
%\caption{The above panel shows the variation of $R (\tau) $ (Fig.~\ref{ScaleFactor}), matter-energy density $\rho_M$ (Fig.~\ref{shell_energy_density}), energy density $\rho_{DE}$ (Fig.~\ref{rhoM})  and the energy density of the shell $\sigma_\Sigma$ with proper time (Fig.~\ref{ShellEnergyDensity}) in the interior spacetime region. Here we have taken the  values of various constants and parameters as follows: $\sin\chi_0 = 1/\sqrt{2}$, $G_0 = 1$, $M = 4.00$, $\alpha = 0.6$.  The initial conditions are chosen such that, at $\tau = 0$, we have  $R = 20$ and $\dot{R} = 0$. For these values of parameters, we have $Q = \sqrt{\alpha G_0} M = 9.6$ and $GM \approx 6.4 $, and consequently $GM > Q$.
}
\label{Dust_IntSol}
\end{figure*}

\begin{figure*}
\begin{subfigure}{.5\textwidth}
  \centering
  \includegraphics[width=1.0\linewidth]
  {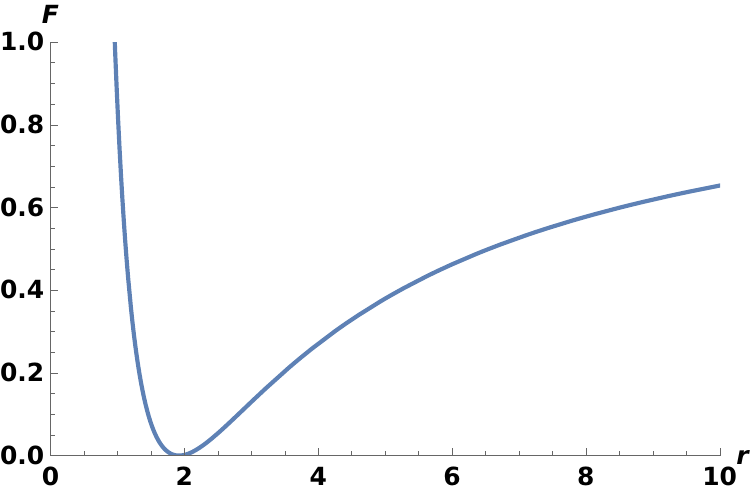}
  \caption{}
  \label{HorizonFunctionm}
\end{subfigure}%
\begin{subfigure}{.5\textwidth}
  \centering
  \includegraphics[width=1.0\linewidth]{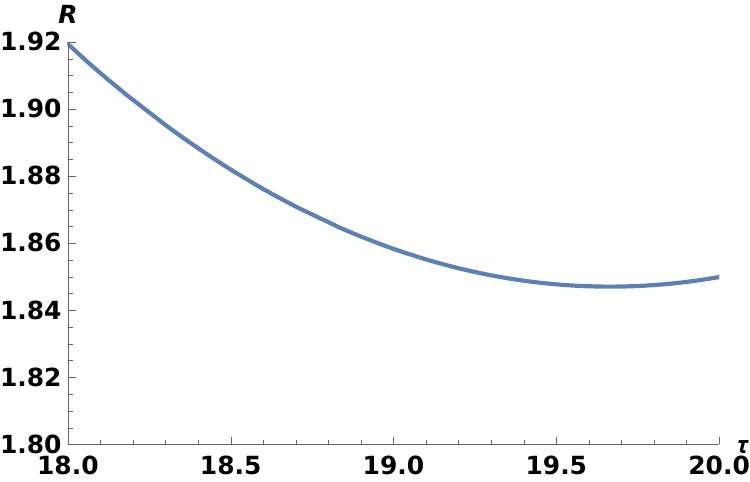}
  \caption{}
  \label{}
\end{subfigure}%
\caption{The above figures show the horizon function. For the parameters we have chosen, the horizon function $F$ becomes zero at $r = 1.92$ and from the image in the right-hand side we see that the boundary of the collapsing matter enters the horizon. 
} 
\label{Dust_IntSol}
\end{figure*}

%\begin{figure*}
%  \centering
%  \includegraphics[width=0.5\linewidth]{Plots/Dust Soln/dust_Horizon_Function.pdf}
%\caption{The above figure shows the variation of the Horizon function w.r.t. the proper time.
%}
%\label{HorizonFunction}
%\end{figure*}

If we assume the shell is composed of baryonic matter and dark energy like component interacting separately with the $A_0(\tau)$ field then: $\sigma_\Sigma=\sigma_{\rm int} + \sigma_{M} + \sigma_{DE}$ and $p_\Sigma=p_{\rm int} + p_{DE}$. Moreover the shell energy-momentum tensor satisfies
\begin{eqnarray}
\dot{\sigma}_\Sigma +  \frac{2\dot{R}}{R}(\sigma_\Sigma + p_\Sigma)= F_{\rm RN}\,,   
\label{emccs}
\end{eqnarray}
where $F_{\rm RN}$ specifies the rate of energy flow to the exterior spacetime, this flow maintains the $A_0(r)$ configuration over an expanding RN-like spacetime. If the interaction of matter and the vector field component on the shell is small compared to the other components then one can indeed have $p_{\Sigma} \sim p_{DE}$ and the energy density component on the shell will be primarily coming from the energy densities of the baryonic part and the dark energy part. 

The resulting collapse is shown in Fig.~[\ref{gen_Dust_IntSol}] and Fig.~[\ref{HorizonFunction}]. In the present paper we have assumed the universal gravitational constant $G_0=1$ in our chosen units. The plots show how the gravitational collapse proceeds in proper time $\tau$. The plots show that as the shell approaches the Cauchy horizon of the RN-like spacetime the shell-energy density rapidly starts to decrease and tends to become negative showing the onset of instabilities related to the Cauchy horizon. Our analytical method fails just when the system has evolved in such a way that the shell is near to the Cauchy horizon. For the asymptotic observer, the STVG-charged shell will never cross the outer horizon of the RN-like black hole. Our calculation shows that in the truncated STVG theory, one can indeed have gravitational collapse leading to black hole formation. 

%%%%%%%%%%%%%%%%%%%%%%%%%%%%%%%%%%%%%%%%%%%%%%%%%%%%%%%%%%%%%%%%%%%%%%%%%%%%%%%%%%%%%%%%%%%%%%%%%%%%%%%%%%%%%%%%%%%%%%
\subsection{Interior solution with Dark Energy, a Dust-like Matter Component and Conformal Shell, where $S=0$ }
%%%%%%%%%%%%%%%%%%%%%%%%%%%%%%%%%%%%%%%%%%%%%%%%%%%%%%%%%%%%%%%%%%%%%%%%%%%%%%%%%%%%%%%%%%%%%%%%%%%%%%%%%%%%%%%%%%%%%%%%%%%%%%%%%%%%%%%%%%%%%%%%%%%%%%%%%%

In this subsection we deal with the case where $G$ is assumed to be a field, with a constant effective value. In this case we do not disregard the equation of motion of $G$ field and consequently it turns out that the energy-momentum tensor of the shell must be traceless, i.e., $S_{ab} h^{ab} = 0$. This implies that, $p_\Sigma = \frac{1}{2} \sigma_\Sigma $, $S^\tau_\tau = - \sigma_\Sigma$ and $S^\vartheta_\vartheta = \frac{1}{2} \sigma_\Sigma  = S^\phi_\phi$. In the present case, the components in the interior spacetime remain the same as before, i.e., a dust-like matter sector and a dark-energy-like fluid and the dynamics for the interior FLRW spacetime is governed by Eqs.~\eqref{RhoM} and \eqref{pM}.

Using the junction condition, Eq.~\eqref{JCShell},  we have the following two equations determining the dynamics of the shell:
\begin{align}
    &\frac{\dot{\beta}}{\dot{R}} = 8 \pi G \sigma_\Sigma ,  \label{shellPressureConf} \\
    &\frac{\beta}{R} - \frac{\cos\chi_0}{R} = -4 \pi G \sigma_\Sigma, \label{shellEnergyConf}
\end{align}
which can be combined to write
\begin{align}
     \frac{\dot{\beta}}{\dot{R}} +
    \frac{2\beta}{R} - \frac{2\cos\chi_0}{R} = 0. \label{DiffEqR}
\end{align}
This  is a second-order differential equation for $R$ which is a function of $\tau$. We now have four unknowns, $R(\tau) $, $\rho_M$,  $\rho_{DE}$, $\sigma_\Sigma$ and four equations (Eqs.~\eqref{RhoM}, \eqref{pM}, \eqref{shellPressureConf}, \eqref{shellEnergyConf}) and consequently the system is closed and can be solved without introducing any other condition.
 Now we can solve Eq.~\eqref{DiffEqR}  to get the solution for $R(\tau)$, which we can use to calculate shell-energy density and matter energy density from Eqs.~\eqref{shellEnergyConf} and \eqref{RhoM}. 
 
 There are some properties of the above mentioned scheme of gravitational collapse. It was seen that the chosen parameter values of the theory and the specific initial conditions can produce a collapse to an extremal RN-like solution where we have $Q^2 = GM$. In this case, both the horizons coincide and the RN-like black hole horizon itself becomes a Cauchy horizon. Moreover, in the present case, the shell energy-momentum tensor is traceless and this condition is restrictive. There can be various possible ways to produce such a condition. The interaction term may suitably modify the other components of the energy density and pressure so that effectively $S = 0$. One may also try to address this issue in a different way. One may assume that the predominant matter field on the shell is the standard photon field that is STVG-charged and is interacting with the STVG 4-vector component on the shell. If these interaction terms are subdominant in the energy-momentum tensor of the shell, then one can indeed have an approximately conformal shell.
 
The results related to the collapse are shown in Fig.~[\ref{Dust_IntSol}] and Fig~[\ref{HorizonFunction}]. In the present case, we have a proper gravitational collapse, and the shell can be traced up to some point inside the Cauchy horizon. In this case, the instabilities may affect the system once it is near the Cauchy horizon or it has just crossed the horizon. Our analytical method shows that the system is near a bounce point as the shell just crosses the Cauchy horizon. To show that the shell actually crosses the horizon, we have presented the nature of the horizon function $F$ and $R$ near the Cauchy horizon in Fig~[\ref{HorizonFunction}], which shows that the shell has indeed crossed the horizon. Once inside the Cauchy horizon in the RN-spacetime, the system is not expected to come out again, assuming that the horizon is stable in the STVG theory. For an asymptotic observer, the charged shell will take an infinite time to cross the horizon.  

%%%%%%%%%%%%%%%%%%%%%%%%%%%%%%%%%%%%%%%%%%%%%%%%%%%%%%%%%%%%%%%%%%%%%%%%%%%%%%%%%%%%%%%%%%%%%%%%%%%%%%%%%%
\section{Discussion and Conclusion}

In this article, we have presented the full junction conditions in STVG theory, which are required when one attempts to glue two different spacetimes along a junction. These conditions are derived in the most general way. Next, we have used the junction conditions in the case of gravitational collapse, where the internal spacetime is like the FLRW spacetime, while the external spacetime is the RN-like spacetime solution in STVG theory. Finally, we have solved the collapse dynamics for two different cases. The first case is a truncated version of the full STVG theory, and the second case captures the complexity of the STVG theory to a certain extent. The results show that in these simplistic toy models, one can indeed have a gravitational collapse of matter, where the end state is an STVG black hole solution.  

The main challenge in working with STVG theory is the presence of  many field degrees of freedom, which complicates any real calculations. To reduce unnecessary complications and still be relevant, we have  simplified the problem considerably in light of previous calculations \cite{Moffat2006, MoffatMOG}. The interior spacetime is practically an FLRW spacetime, as one has it in GR, and throughout, we have frozen the scalar degrees of freedom. The resulting theory is an STVG theory where the tensor and vector components are the primary components of the theory. 

The present paper presents two simplified collapse solutions. The first simplification is related to the interaction terms. In this paper, we did not explore the interaction Lagrangian, which dictates the coupling between matter and vector components of the theory. The primary reason for this is that the interaction Lagrangian landscape is a rich and complicated one, and consequently, it requires a separate dedicated investigation. On the whole, whenever there is such a coupling, it may not be producing energy density and pressure comparable to the corresponding values for normal baryonic matter or dark energy component. 
%The reason is that if the interaction terms contribute more to the energy-momentum tensor, the identities of the interacting %components become ambiguous. 
%In this light, one may expect that the matter-4-vector interaction energy densities may play a subdominant role on the shell %separating the two spacetimes. 
We assume that the non-minimal interaction between the shell matter and the STVG vector component $A_0$ is sufficiently weak that its contribution to the surface energy-momentum tensor is subdominant compared with the intrinsic matter and dark-energy components. The interaction is nevertheless responsible for generating the effective STVG surface current which sources the exterior RN-like vector field. This assumption preserves the physical identity of the shell constituents while allowing the shell to carry STVG charge. The other simplification is related to the instabilities of the shell when it nears the Cauchy horizon. As specified in the introduction, the Cauchy horizon sets the limit of predictable gravitational dynamics, which starts from initial conditions given outside of the Cauchy horizon. When the shell nears such a Cauchy horizon, there can be various instabilities. In the present article, we did not discuss the issues related to these instabilities.  

The choice of coordinates plays an interesting part in the collapsing models presented. In the whole paper we have used Schwarzschild/Reissner--Nordstr\"{o}m coordinates. This coordinate system makes the problem simpler at the cost of some unphysical singularities. As an example if one assumes the matter-field interaction \cite{Moffat:2013sja} on the shell to be of the form $u^0 A_0$ then $u^0 A_0 \sim 1/F$ where $F$ is the horizon function and attains zero value whenever the shell crosses the  RN-like horizons. As a result the interaction term may diverge at the instant when the shell crosses the horizons. One need not be alarmed at such a divergence as this is a coordinate effect and physically no fields diverge on the external horizon of the RN-like black hole. In our paper we have practically assumed the interaction energy-momentum tensor components to be negligible compared to the other contributions in $S_{\mu \nu}$ and consequently this possible unphysical divergence did not show up. This scheme of avoiding the possible unphysical singularity of the interaction term on the electrically charged shell is not new, while discussing shell collapse in Einstein-Maxwell theory in GR, the author of Ref.~\cite{kuch} used a similar idea.  

In STVG theory, both the FLRW-like spacetime and RN-like spacetime can accommodate a finite $A_0$ component of the 4-vector field. It is seen that in such a case, if we try to match the $A_0$ field continuously over the junction, we encounter a particular kind of singularity when the junction crosses the outer horizon of the RN-like spacetime. Such a singularity is not physical, it is a coordinate singularity. Such a singularity can be averted, if we match the $B_{\mu\nu}$ field on the junction. This kind of matching is natural in a Maxwell-like 4-vector theory where the gauge field matching at the junction is not recommended as the condition can be gauge dependent. In the present case, we have a Proca-like theory for the 4-vectors. By assuming that the 4-vector field is massless in both spacetimes, we have transformed the Proca-like theory into an effective Maxwell-like theory for the 4-vector field. Consequently, our matching conditions are consistent.  

One of the predictions from our work is that gravitational collapse of the FLRW core glued to an RN-like spacetime outside must happen through an STVG-charged shell, which produces the 4-vector field component in the external spacetime. The novelty of STVG theory is that it forces the shell to have a traceless energy-momentum tensor. The tracelessness of $S_{\mu \nu}$ is restrictive, and is a distinctive feature of the STVG theory where $G$ is a scalar field, but can be realized if one assumes electromagnetic photons to be distributed over the shell, which are nominally interacting with the STVG 4-vector field. In that case, the interactions will not be able to affect the properties of the energy-momentum tensor for the photons on the shell substantively and one may have the traceless energy-momentum tensor on the shell in an approximate way.

After the work of Oppenheimer and Snyder on gravitational collapse, several other authors studied the same with charged matter in the framework of GR ~\cite{kuch, beken, vick, kehle}. In \cite{mashhoon} authors extensively studied  the gravitational collapse of charged fluid sphere and as a special case worked out uniform density model. In astrophysics we rarely find charged compact objects, generally compact objects are electrically neutral. An electrically neutral, non-rotating black hole in GR is the Schwarzschild black hole. In the present model of gravitational collapse in STVG theory it is seen that the resulting black hole may not be a Schwarzschild like black hole. Observational tests involving lensing of light and black hole shadow formation will yield results which differ from an ideal Schwarzschild black hole. If we can verify that the black hole is electrically uncharged by some other observations, such signals will predict the existence of STVG black hole solutions. In this article we have given a dynamical collapse mechanism in which such exotic STVG black holes can be produced.   

%%%%%%%%%%%%%%%%%%%%%%%%%%%%%%%%%%%%%%%%%%%%%%%%%%%%%%%%%%%%%%%%%%%%%%%%%%%%%%%%%%%%%%%%%%%%%%5
\section{Acknowledgments}

Debanjan Debnath expresses his sincere gratitude to the Council of Scientific and Industrial Research (CSIR, India, CSIR Award No.: CSIRAWARD/SPM-JR2022/13184) for funding the work.
\end{document}